\documentclass[superscriptaddress,nofootinbib,a4paper,11pt]{article}
\usepackage{jheppub}
\usepackage{graphicx}
\usepackage{subcaption}
\usepackage{float}
\usepackage{amsmath}
\usepackage{amssymb}
\usepackage[utf8]{inputenc}
\usepackage{hyperref}
\usepackage{color}
\usepackage{slashed}
\usepackage{multirow}
\usepackage{enumitem}
\usepackage{lineno}
\usepackage{booktabs}
\usepackage{multirow} 
\usepackage{siunitx}  
\usepackage{threeparttable}
\usepackage{changes}
\allowdisplaybreaks

\begin{document}

\title{\boldmath Search for Cosmic-Ray Produced Dark Meson via the $U(1)_\text{D}$ Portal at JUNO}

\author[a]{Zirong Chen,}
\author[b]{Dan Chi,}
\author[a]{Jinmian Li,}
\author[c]{Junle Pei}

\affiliation[a]{College of Physics, Sichuan University, Chengdu 610065, China}
\affiliation[b]{Provincial Key Laboratory of Solid-State Optoelectronic Devices, Zhejiang Normal University, Jinhua, 321004, China}
\affiliation[c]{Institute of Physics, Henan Academy of Sciences, Zhengzhou 450046, China}

\emailAdd{chenzirong@stu.scu.edu.cn}
\emailAdd{chidan@zjnu.edu.cn}
\emailAdd{jmli@scu.edu.cn}
\emailAdd{peijunle@hnas.ac.cn}

\abstract{
We investigate the atmospheric production and subsequent detection of sub-GeV dark mesons within the framework of a confining dark sector coupled to the Standard Model (SM) via a $U(1)_{\text{D}}$ vector portal. 
High-energy cosmic ray interactions in the atmosphere produce dark quarks through proton bremsstrahlung, rare decays of Standard Model mesons, and Drell-Yan processes, which subsequently hadronize into dark mesons. 
We adopt a modified Quark Combination Model to describe the non-perturbative dark hadronization process, allowing for a detailed event-level characterization of the dark meson flux. 
We simulate the flux and the interactions of these relativistic dark mesons in the Jiangmen Underground Neutrino Observatory (JUNO) using the GENIE generator, considering both elastic scattering off nuclei and deep inelastic scattering channels. 
Adopting a LES-inspired conservative signal definition with $15~\mathrm{MeV}<E_{\rm vis}<100~\mathrm{MeV}$ and vetoing events containing final-state neutrons, we derive projected $90\%$ C.L. sensitivities to the coupling strength between the dark gauge boson and the SM sector for a $20~\mathrm{kton}\!\cdot\!\mathrm{year}$ exposure. For the representative benchmark masses $m_{Z^\prime}=0.001$, $0.01$, and $0.1~\mathrm{GeV}$, the projected JUNO reach corresponds to $g_{\rm SM}^{90}=6.48\times10^{-5}$, $7.75\times10^{-5}$, and $2.37\times10^{-4}$, respectively. This reach is found to be only mildly sensitive to order-one variations of the hadronization parameter, as illustrated by the choices $\kappa_\beta=1/3,\,1,$ and $3$.}

\maketitle

\section{Introduction} \label{sec:1} 

The Standard Model (SM) of particle physics has achieved remarkable success in describing fundamental interactions. However, it fails to explain the existence of Dark Matter (DM).
While Weakly Interacting Massive Particles (WIMPs) have long been the leading candidates, the null results from direct detection experiments and the Large Hadron Collider (LHC) have motivated the exploration of alternative paradigms, particularly those involving light dark sectors in the sub-GeV mass range~\cite{Battaglieri:2017aum}. A theoretically well-motivated class of such models involves a ``Hidden Valley" or a strongly interacting dark sector, characterized by a new non-Abelian gauge group $SU_{\text{D}}(N)$~\cite{Strassler:2006im,Han:2007ae,Kribs:2016cew,Gori:2022vri}. Analogous to Quantum Chromodynamics (QCD), this dark sector confines at low energies, giving rise to a rich spectrum of dark hadrons. Among those bound states, the lightest dark hadrons, which are often identified as pseudo-Nambu-Goldstone bosons or ``dark pions" ($\pi_D$)~\cite{Bai:2010qg,Buckley:2012ky,Cline:2013zca,Harigaya:2016rwr,Beylin:2019gtw,Beylin:2020bsz,Alexander:2023wgk,Chung:2025wle}, can be stable and serve as viable dark matter candidates. This scenario naturally realizes the Strongly Interacting Massive Particle (SIMP) mechanism~\cite{Hochberg:2014dra,Hochberg:2014kqa,Lee:2015gsa,Hochberg:2015vrg}, which not only provides the correct relic abundance but also offers a potential solution to the small-scale structure problem~\cite{Bullock:2017xww,Tulin:2017ara}. 

The phenomenology of these strongly interacting dark sectors heavily depends on the portal connecting them to the visible sector. A minimal and predictive scenario is established via the kinetic mixing between a new $U(1)_{\text{D}}$ gauge boson and the SM photon~\cite{Holdom:1985ag,Foot:1991kb}. Through this vector portal, dark quarks can be produced in high-energy collisions, subsequently hadronizing into dark mesons. 
Extensive constraints on such scenarios have been placed by accelerator-based beam dump experiments and collider searches, which target signatures such as displaced vertices induced by the long-lived dark meson~\cite{Cheng:2021kjg,Born:2023vll,Cheng:2024hvq,Cheng:2024aco,Liebersbach:2024kzc,Liu:2025bbc,CMS:2025mym,Bernreuther:2025xqk}, semi-visible jets~\cite{Cohen:2015toa,Beauchesne:2017yhh,CMS:2021dzg,Liu:2024rbe,Buckley:2025hty,ATLAS:2025kuz}, emerging jets~\cite{Schwaller:2015gea,CMS:2018bvr,Carrasco:2023loy,ATLAS:2025lfx,Carrasco:2025bct} and so on. 
Complementary to terrestrial accelerators, high energy cosmic rays colliding with the Earth’s atmosphere (also dubbed as cosmic-ray air shower) provide a complementary and permanent source of sub-GeV dark sector particles~\cite{Kusenko:2004qc,Yin:2009yt,Arguelles:2019ziu,Coloma:2019htx,Alvey:2019zaa,Su:2020zny,Cheung:2022umw,Airen:2025uhy}. This atmospheric production mechanism has been extensively studied for Millicharged Particles (MCPs)~\cite{Hu:2016xas,Plestid:2020kdm,ArguellesDelgado:2021lek,Kachelriess:2021man,Du:2022hms,Wu:2024iqm}, but its application to composite dark sectors remains less explored due to the complexities arising from the non-perturbative dark hadronization process. 

In this work, we investigate the atmospheric production and subsequent detection of sub-GeV scale dark mesons within the framework of a dark $SU(N)$ gauge theory coupled to the SM via a $U(1)_{\text{D}}$ portal. We will focus on the coupling between the new gauge boson and the SM quarks~\cite{Tulin:2014tya,Batell:2014yra,Soper:2014ska,Dobrescu:2014ita,Kouvaris:2016afs}. The leptonic couplings can be assumed to be suppressed to evade the stringent bounds from electron beam dump and lepton collider experiments. 
We construct the low-energy effective Lagrangian based on Chiral Perturbation Theory (ChPT) to describe the dynamics of the dark pions. The flux of dark mesons originating from cosmic-ray proton interactions is calculated by considering three distinct production channels: proton bremsstrahlung, the decay of SM mesons (specifically $\pi^0$, $\eta$) into dark gauge boson, and the Drell-Yan production of dark quarks.

A critical challenge in modeling the flux of composite dark matter is the description of the hadronization process from dark quarks to dark mesons. Unlike previous studies that often rely on phenomenological hadronization models or generic scaling of \textsc{Pythia} parameters~\cite{Carloni:2010tw,Carloni:2011kk} to approximate a confining $SU(N)$ gauge interaction, we employ a Modified Quark Combination Model (MQCM)~\cite{Xie:1988wi,Wang:1996jy,Si:1997zs,Si:1997rp,Si:1997ux,Han:2007ae} to describe the hadronization process. 
Under the assumption of an approximate flavor symmetry within the dark sector, this framework enables a detailed, event-by-event characterization of the multiplicity and kinematics of dark mesons generated in atmospheric dark showers.

Once produced, these relativistic dark mesons travel through the Earth and can scatter off electrons or nuclei in deep underground detectors. The Jiangmen Underground Neutrino Observatory (JUNO)~\cite{JUNO:2015zny,JUNO:2021vlw,JUNO:2025gmd}, with its large fiducial mass (20 kton) and excellent energy resolution, offers an ideal environment for probing such light, weakly interacting particles. We calculate the scattering rates of the atmospheric dark mesons flux within the JUNO detector utilizing the boosted dark matter module~\cite{Berger:2018urf,Berger:2025zfs} implemented in the \textsc{GENIE} framework~\cite{Andreopoulos:2009rq,Andreopoulos:2015wxa}, thereby deriving projected sensitivity limits on the model parameter space. To obtain the projected sensitivity in a background-controlled region, we adopt a Large-Energy-Singles (LES)--inspired conservative signal definition, requiring $15~\mathrm{MeV}<E_{\rm vis}<100~\mathrm{MeV}$ and vetoing events containing final-state neutrons, broadly following the strategy proposed for JUNO in Ref.~\cite{Chauhan:2021fzu}.

The remainder of this paper is organized as follows. In Section~\ref{sec:2}, we outline the theoretical framework of the dark $SU(N)$ model and construct the effective Lagrangian for dark pions. 
Section~\ref{sec:3} elaborates on the implementation of the MQCM.
Section~\ref{sec:4} details the calculation of the atmospheric dark meson flux, incorporating contributions from proton bremsstrahlung, SM meson decay, and Drell-Yan production. 
In Section~\ref{sec:5}, we describe the evaluation of signal event rate at JUNO and present the projected sensitivities. Finally, we summarize our findings and discuss their implications in Section~\ref{sec:6}.

\section{The Dark Matter Model} \label{sec:2}

Analogous to the QCD in the SM, we postulate a dark sector governed by a non-Abelian $SU(N)_{\text{D}}$ gauge interaction, which exhibits both asymptotic freedom and dark quark confinement. The Lagrangian density for dark quarks and dark gluons is given by:
\begin{align}
    \mathcal{L}_{\mathrm{dQCD}} = -\frac{1}{4} G_{\mu \nu}^a G_a^{\mu \nu} + \sum_{k,j=1}^{N_f} \bar{q}_k(i\slashed{D}\delta_{k,j} - m_{k,j}) q_j~, \label{lag}
\end{align}
where $G_{\mu \nu}^a$ ($a = 1, 2, \ldots, N^2 - 1$) denotes the field strength tensor of the dark gluons, and $M_q = (m_{k,j})$ is the mass matrix for the $N_f$ dark quark flavors. For simplicity, we assume a diagonal mass matrix:
\begin{align}
    M_q = \text{diag}(m_1, m_2, \ldots, m_{N_f})~. \label{mass}
\end{align}

\subsection{The $U(1)_{\text{D}}$ Portal Interaction}

We introduce a dark $U(1)_{\text{D}}$ gauge symmetry that mediates interactions between the dark quarks and the SM quark. The interaction Lagrangian for the dark quarks is:
\begin{align}
    \mathcal{L}_{\mathrm{int}} \supset -g^\prime Z^\prime_\mu \sum_{k=1}^{N_f} \bar{q}_k \gamma^\mu (Q^\prime_{k,L} P_L + Q^\prime_{k,R} P_R) q_k~, \label{lagdu1}
\end{align}
where $Z^\prime_\mu$ and $g^\prime$ represent the dark gauge field and its associated gauge coupling, respectively. The parameters $Q^\prime_{k,L/R}$ denote the $U(1)_{\text{D}}$ charges of the left- and right-handed dark quarks $q_{k,L/R}$.

To establish a portal to the SM, the $Z^\prime$ boson couples to the SM quarks as follows:
\begin{align}
\mathcal{L}_{\rm int} \supset -g_{\rm SM} Z^\prime_\mu \bar{\psi}_f \gamma^\mu \psi_f~,
\end{align}
where $f = \{u, d, s, c\}$ denotes the light SM quark flavors relevant to our analysis. For simplicity, the $U(1)_{\text{D}}$ charges of SM fermions are absorbed into the effective coupling $g_{\rm SM}$. We assume purely vector-like couplings to the SM and require the $Z^\prime$ to be lepton-phobic to evade stringent constraints from lepton beam dump and collider experiments.

\subsection{Dark Mesons and Chiral Symmetry Breaking} \label{sec:meson}

In the case where dark quark masses are zero, the Lagrangian density in Eq.~(\ref{lag}) exhibits independent chiral symmetry between the left-handed and right-handed quark fields, enabling independent \( SU(N_f)_L \) and \( SU(N_f)_R \) transformations. It also possesses \( U(1)_V \) symmetry for global phase transformations and \( U(1)_A \) symmetry for axial transformations, the latter at a classical level. The global symmetry is described by
\begin{align}
    U(N_f)_L \times U(N_f)_R = SU(N_f)_L \times SU(N_f)_R \times U(1)_V \times U(1)_A~.
\end{align}
When a non-zero dark-quark condensate \(\langle \bar{q} q \rangle\) forms at an energy scale below the confinement scale \(\Lambda_{\text{dQCD}}\), the chiral symmetry \( SU(N_f)_L \times SU(N_f)_R \) undergoes spontaneous symmetry breaking to a vector symmetry \( SU(N_f)_V \),
\begin{align}
    SU(N_f)_L \times SU(N_f)_R \rightarrow SU(N_f)_V~.\label{viol}
\end{align}
According to the pattern of spontaneous symmetry breaking in Eq.~(\ref{viol}), Goldstone bosons, identified as dark mesons \(\pi_a\) (\(a = 1, 2, \ldots, N_f^2 - 1\)), are generated. These dark mesons remain massless in the limit where dark quark masses are absent. However, due to the explicit breaking of chiral symmetry by the dark quark masses in Eq.~(\ref{mass}), the dark mesons acquire small masses.

At the lowest order, these dark mesons can be effectively described by the \(\sigma\) model, with the Lagrangian density given by
\begin{align}
    \mathcal{L}_{\mathrm{eff}} = \frac{F^2_0}{4} \operatorname{Tr}[(D_\mu U)(D^\mu U)^{\dagger}] + \frac{F_0^2 B_0}{2} \operatorname{Tr}[M_q U^\dagger + U M^\dagger_q]~,\label{mesonmass}
\end{align}
where
\begin{align}
   U = e^{\frac{2i}{F_0}\pi_a T^a}~.
\end{align}
In this expression, \(F_0\) represents the decay constant of the dark mesons, and \(T^a\) (\(a = 1, 2, \ldots, N_f^2 - 1\)) are the generators of the \(SU(N_f)\) group.

Based on the chirality of the fermions, we represent the dark quark fields as
\begin{align}
    q_k=\left(
    \begin{array}{c}
    q_{k,L} \\
    q_{k,R} \\
\end{array}
    \right).
\end{align}
We define
\begin{align}
    2\pi_a T^a+\sqrt{\frac{2}{N_f}}\eta_1 I_{N_f \times N_f}=\sqrt{2} \pi^\prime ~,
\end{align}
where \(I_{N_f \times N_f}\) is the \(N_f \times N_f\) identity matrix. The correspondence between the matrix elements of \(\pi^\prime\) and the dark quark fields is given by
\begin{align}
   & \pi^\prime_{k,j}\longleftrightarrow i \bar{q}_k\gamma^5 q_j=i \left(q_{k,L}^\dagger q_{j,R}-q_{k,R}^\dagger q_{j,L}\right).
\end{align}
Under the transformations of parity (\(\hat{P}\)) and charge conjugation (\(\hat{C}\)), these fields satisfy
\begin{align}
   & \hat{P}\pi^\prime_{k,j} \hat{P}=-\pi^\prime_{k,j}~,\\
   & \hat{C}\pi^\prime_{k,j}\hat{C}=\pi^\prime_{j,k}~.
\end{align}
In conclusion, these dark mesons exhibit pseudoscalar characteristics, as evidenced by their transformations under parity and charge conjugation.

The matrix elements of \(\pi^\prime\) or the combinations of these elements lead to \(N_f^2 - 1\) dark mesons \(\pi_a\) and a singlet meson \(\eta_1\). When \(k \neq j\), \(\pi^\prime_{k,j}\) corresponds to complex neutral and charged dark mesons, totaling \(N_f^2 - N_f\). For instance, in QCD with \(N_f = 3\), we find \(\pi^\prime_{1,2/2,1} = \pi^\pm\), \(\pi^\prime_{1,3/3,1} = K^\pm\), and \(\pi^\prime_{2,3/3,2} = K^0/\bar{K}^0\). According to Eq.~(\ref{mesonmass}), we have the squared masses of $\pi^\prime_{k,j}$ as
\begin{align}
    M^2_{\pi^\prime_{k,j}}=B_0 (m_k+m_j)~,~~~k\neq j~.
\end{align}
Among \(\pi_a~(a=1,2,\ldots,N_f^2-1)\), we denote the real neutral dark mesons as \(\pi_A~(A=1,2,\ldots,N_f-1)\), which are related to \(\pi^\prime_{k,k}~(k=1,2,\ldots,N_f)\) by
\begin{align}
\pi_A=\sqrt{2}\sum_{k=1}^{N_f}\left(T^A_{k,k}\pi^\prime_{k,k}\right).\label{piA}
\end{align}
In this, \(T^A\) are the diagonal generators of the \(SU(N_f)\) group corresponding to \(\pi_A\), expressed as
\begin{align}
    T^A_{k,j}=\frac{\delta_{k,j}}{\sqrt{2A(A+1)}}\times \begin{cases}
1~,\quad & 1\leq k< A+1 \\
-A~,\quad &k=A+1 \\
0~,\quad & A+1<k\leq N_f 
\end{cases}~,~~~A=1,2,\ldots,N_f-1~. 
\end{align}
Equation (\ref{mesonmass}) gives
\begin{align}
    \mathcal{L}_{\mathrm{eff}}\supset -\frac{1}{2}\sum_{A,B=1}^{N_f-1} \left[M^2_\pi\right]_{A,B}\pi_A \pi_B~,
\end{align}
where
\begin{align}
  \left[M^2_\pi\right]_{A,B}=  4 B_0 \text{Tr}\left[T^A T^B M_q\right]~.
\end{align}
Quark compositions and squared masses of dark mesons are summarized in Table~\ref{tab:mass}.
In the scenario where all dark quarks possess identical masses, i.e.,
\begin{align}
    m_k = m_q~, \quad k = 1, 2, \ldots, N_f~, \label{mall}
\end{align}
all dark mesons will consequently exhibit the same mass, given by
\begin{align}
    M_{\pi^\prime_{k,j}} = M_{\pi_A} = \sqrt{2B_0 m_q}, \quad k, j = 1, 2, \ldots, N_f \text{ with } k \neq j, \quad A = 1, 2, \ldots, N_f - 1~. \label{piall}
\end{align}

\begin{table}[htb]
	\centering
	\begin{tabular}{ccc}  
		\hline
		dark meson &   quark composition  &  squared mass \\
		\hline 
             $\pi^\prime_{k,j}~(k \neq j)$ & $i\bar{q}_k\gamma^5 q_j$ & $M^2_{\pi^\prime_{k,j}}=B_0 (m_k+m_j)$ \\~\\
             $\pi_A~(A=1,2,\ldots,N_f-1)$ &  $i\sqrt{2} \sum_{k=1}^{N_f}\left(T^A_{k,k} \bar{q}_k\gamma^5 q_k\right)$ & $\left[M^2_\pi\right]_{A,B}=4 B_0 \text{Tr}\left[T^A T^B M_q\right]$ \\~\\
              $\eta_1$ & $\frac{i}{\sqrt{N_f}}\sum_{k=1}^{N_f}\left(\bar{q}_k\gamma^5 q_k\right)$ &  \\
		\hline
	\end{tabular}
	\caption{\label{tab:mass} Quark compositions and squared masses of dark mesons.} 
\end{table}

The parameters mentioned above are related approximately by
\begin{align}
    &F_0\sim \Lambda_{\text{dQCD}}~,\\
    &N_f F_0^2 B_0=-\langle \bar{q} q \rangle~.
\end{align}
Thus, in the numerical calculations performed in this study, we choose \(F_0\), \(B_0\), and the dark quark masses \(m_k\) (\(k = 1, 2, \ldots, N_f\)) as the fundamental parameters.

In Eq.~(\ref{mesonmass}), the covariant derivative is expressed as
\begin{align}
    D_\mu U=\partial_\mu+i U\cdot L_\mu-i R_\mu\cdot U 
\end{align}
with
\begin{align}
   & L_{\mu}=-g^\prime Z^\prime_\mu \text{diag}\left(Q^\prime_{1,L},Q^\prime_{2,L},\ldots,Q^\prime_{N_f,L}\right)~,\\
    & R_{\mu}=-g^\prime Z^\prime_\mu \text{diag}\left(Q^\prime_{1,R},Q^\prime_{2,R},\ldots,Q^\prime_{N_f,R}\right)~.
\end{align}
Then, the first term in Eq.~(\ref{mesonmass}) gives the interaction between dark mesons and the $U(1)_{\text{D}}$ gauge boson, which contains
\begin{align}
   \mathcal{L}_{\text{eff}}\supset &  g^\prime F_0 Z^\prime_\mu \sum_{A=1}^{N_f-1} \left(\sum_{k=1}^{N_f}T^A_{k,k}\left(Q^\prime_{k,R}-Q^\prime_{k,L}\right)\right) \partial^\mu \pi_A   \nonumber\\
   &-\frac{i}{2}g^\prime Z^\prime_\mu \sum_{1\le k<j\le N_f}   \left(Q^\prime_{k,L}+Q^\prime_{k,R}-Q^\prime_{j,L}-Q^\prime_{j,R}\right) \left(\pi^\prime_{j,k} \partial^\mu\pi^\prime_{k,j}-\pi^\prime_{k,j} \partial^\mu\pi^\prime_{j,k}\right)~. \label{eq:zkk}
\end{align}

As a benchmark scenario for the following study, we assume $N_f = 3$ dark quark flavors with nearly degenerate masses. This leads to a spectrum of dark mesons where the Kaon-like and Pion-like states are the lightest degrees of freedom, with masses determined by the chiral symmetry breaking scale and quark masses. Due to the near-degeneracy, the kinematic distributions of dark mesons produced via the decays of heavier resonances remain largely insensitive to the small mass splittings. Furthermore, we assume vector-like $U(1)_{\text{D}}$ charges for the dark quarks, specifically setting $Q^\prime_{k,L} = Q^\prime_{k,R} = 1$ for two flavors ($k=1,2$) and $Q^\prime_{3,L} = Q^\prime_{3,R} = -1$ for the third, ensuring a simplified yet phenomenologically rich portal structure.
Under this assignment, the $Z^\prime$ boson specifically couples to four flavor-carrying dark mesons (the Kaon-like states $\pi^\prime_{13}$ and $\pi^\prime_{23}$), {while the remaining mesons, including the three pion-like states and two diagonal eta-like mesons}, remain uncoupled to the $Z^\prime$ due to the vector-like nature of the interaction and the charge degeneracy of the first two flavors.
{Analogous to the situation in the Standard Model, we allow for mixing between the octet state $\pi_{a=8}$ (equivalently, $\pi_{A=2}$) and the singlet state $\eta_1$, with the resulting mass eigenstates denoted by $\eta_D$ and $\eta^\prime_D$, respectively, where $\eta^\prime_D$ is more singlet-like and heavier than $\eta_D$.
}
In what follows, for simplicity, $\pi_D$ shall denote a generic dark meson, while $K_D$ is reserved exclusively for those that couple to the $Z^\prime$.

\section{From Quark to Meson: modified Quark Combination Model}
\label{sec:3}

In the previous section, we summarized the effective low-energy description of the light dark-meson spectrum and couplings. To compute fluxes and detector event rates, however, an event-level prescription is required to map the initially produced dark quark pair into a set of dark meson four-momenta. Since hadronization is intrinsically non-perturbative, we model this stage using an MQCM, implementing the QCM plus longitudinal phase space approximation (LPSA) procedure adopted in Ref.~\cite{Han:2007ae}. In this setup, the final-state multiplicity and kinematics are generated stochastically, fixed by energy--momentum conservation in the hadronization rest frame. The resulting events are subsequently boosted to the laboratory frame to serve as input for propagation and scattering-rate calculations.


In our setup, the dark quark pair is produced through the $Z^\prime$ mediation.
The relevant energy scale for hadronization is the invariant mass of the dark-quark system,
\begin{equation}
\sqrt{s}\equiv m_{q_k\bar q_k}=\sqrt{(p^\mu_{q_k}+p^\mu_{\bar q_k})^2}\,.
\end{equation}
For an on-shell $Z'$ production and decay, one simply has $p^\mu_{Z'}=p^\mu_{q_k}+p^\mu_{\bar q_k}$, implying $\sqrt{s}=m_{Z'}$. We generate the final-state momenta in the $Z^\prime$ rest frame and then boost the full event to the laboratory frame using the (off-shell) $Z'$ kinematics determined in the production stage.

For concreteness, we denote the meson as $\pi_D$ and assume all mesons have nearly degenerate mass $m_{\pi_D}$. 
Following Ref.~\cite{Han:2007ae}, we introduce the constituent-mass parameter entering the MQCM multiplicity ansatz,
\begin{equation}
m_{q_k} =\frac{m_{\pi_D}}{2}\,,
\end{equation}
which acts as a phenomenological parameter internal to the hadronization model.

Within the quark-combination picture, the total number of dark mesons produced in an event, $N$, is assumed to obey a shifted-Poisson law:
\begin{equation}
P(N)=\frac{\langle N\rangle^{\,N-1}}{(N-1)!}\,e^{-\langle N\rangle}\,,
\end{equation}
where the shift by $1$ accounts for the original $q_k\bar q_k$ pair. The average meson multiplicity is parameterized as~\cite{Han:2007ae}:
\begin{equation}
\langle N\rangle
=
\sqrt{\alpha^2+\beta\sqrt{s}}-\alpha-1,
\qquad
\alpha=\beta m_{q_k}-\frac14\,.
\label{eq:meanN}
\end{equation}
The parameter $\beta$ governs the multiplicity scaling.
In the QCM literature, $\beta$ is treated as a phenomenological free parameter, and a commonly used benchmark choice in SM QCD applications is $\beta_{\rm BM}=4.2~\mathrm{GeV}^{-1}$, calibrated to $e^+e^-$ annihilation data~\cite{Xie:1988wi,Si:1997ux,Han:2007ae}. In the present dark-sector setup, however, there is no first-principles determination of the corresponding parameter for a generic confinement scale $\Lambda_D$. We therefore adopt the following benchmark parametrization for $\beta$, with $\Lambda_D \equiv m_{\pi_D}$:
\begin{equation}
\beta(\Lambda_D)
=
\beta_{\rm BM}\,
\Bigg[1+\Big(\frac{1~\text{GeV}}{\Lambda_D}\Big)^{p}\Bigg]^{1/p}~.
\label{eq:betaScaling}
\end{equation}

Numerical checks indicate that the final observables are only weakly sensitive to the precise value of the exponent $p$; throughout this work we therefore fix $p=2$ for definiteness. Equation~\eqref{eq:betaScaling} should be viewed as a phenomenological benchmark ansatz for exploring how the average dark-meson multiplicity may vary with the confinement scale, rather than as a matching condition requiring $\beta(\Lambda_D=\Lambda_{\rm QCD})=\beta_{\rm BM}$. For the robustness study, we introduce a dimensionless rescaling factor $\kappa_\beta$ via $\beta(\Lambda_D)\to \kappa_\beta\,\beta(\Lambda_D)$, and consider the representative choices $\kappa_\beta=1/3,\,1,$ and $3$.
Its role is to provide a simple confinement-scale dependence in the event-level hadronization prescription used to map the partonic configuration into dark-meson spectra.

\begin{figure}[t]
    \centering
    \includegraphics[width=0.8\textwidth]{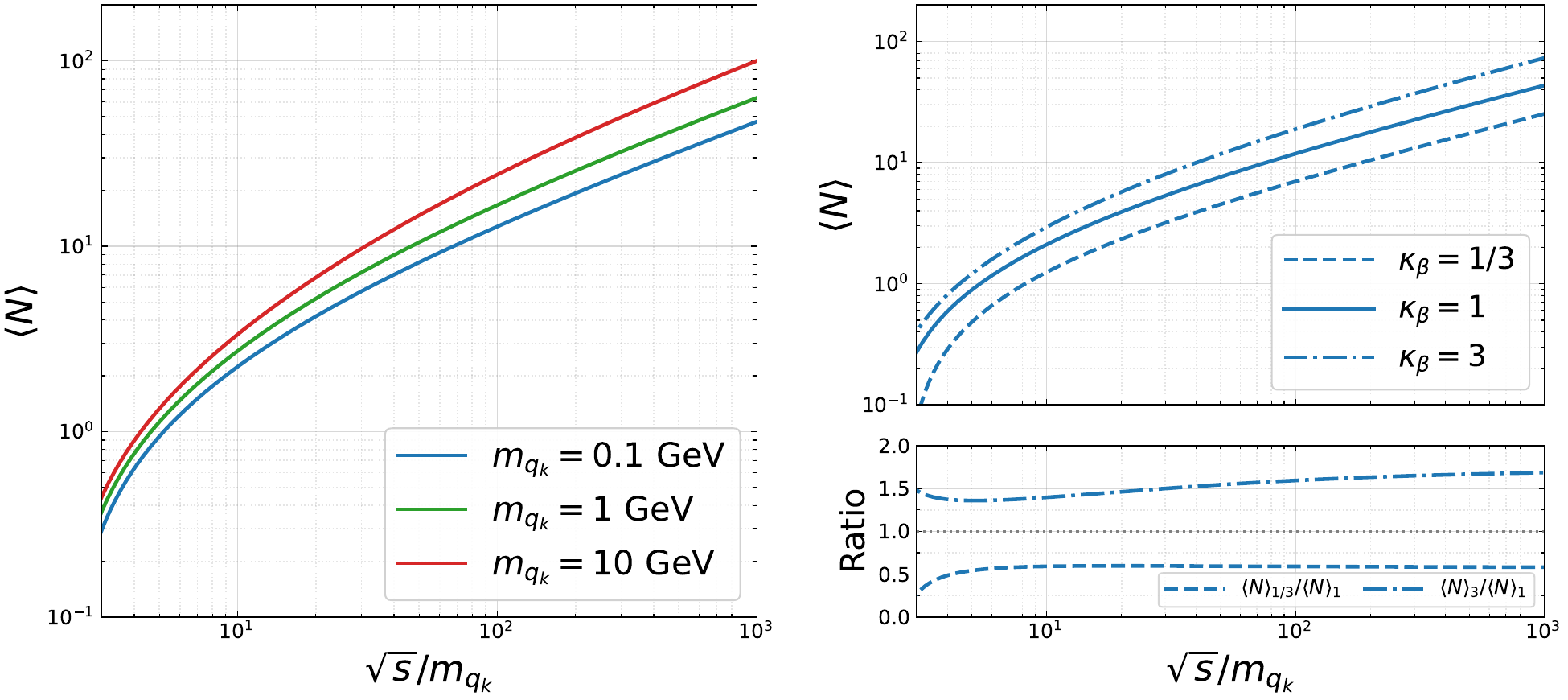}
    \caption{
        Left: average meson multiplicity $\langle N\rangle$ as a function of the dimensionless scaled center-of-mass energy $\sqrt{s}/m_{q_k}$ for three benchmark dark-quark masses, evaluated with $\kappa_\beta=1$.
        Right: average meson multiplicity under $\mathcal{O}(1)$ variations of $\beta$ ($m_{q_k}=0.1~\mathrm{GeV}$). Upper panel: $\langle N\rangle$ for $\kappa_\beta = 1/3, 1, 3$; lower panel: ratios $\langle N\rangle_{\kappa_\beta}/\langle N\rangle_{\kappa_\beta=1}$.     
    }
    \label{fig:multiplicity_scaling}
\end{figure}

To illustrate both the behavior implied by Eq.~\eqref{eq:betaScaling} and its sensitivity to an order-one variation of the multiplicity parameter, we show the resulting $\langle N\rangle$ distributions in Figure~\ref{fig:multiplicity_scaling}. In the left panel, the average meson multiplicity $\langle N\rangle$ is shown for three benchmark dark-quark masses with the $\kappa_\beta=1$. As expected, $\langle N\rangle$ increases with the available phase space. At fixed $\sqrt{s}/m_{q_k}$, the curves do not collapse onto a single universal trajectory, indicating that in the present phenomenological setup the multiplicity depends not only on the kinematic ratio $\sqrt{s}/m_{q_k}$ but also on the absolute scale entering the ansatz for $\beta(\Lambda_D)$. 
In the right panel, we illustrate the robustness of this prescription under the rescaling $\beta\to \kappa_\beta \beta$ for a representative benchmark $m_{q_k}=0.1~\mathrm{GeV}$. The upper panel shows that varying $\kappa_\beta$ mainly shifts the overall normalization of $\langle N\rangle$, while preserving its qualitative dependence on $\sqrt{s}/m_{q_k}$. 
The lower panel uses the ratios $\langle N\rangle_{\kappa_\beta}/\langle N\rangle_{\kappa_\beta=1}$ to show the deviation from the fiducial result. The resulting variation in $\langle N\rangle$ is of similar size to the value of the $\beta$ parameter.

To construct the four-momenta of the $N$ dark mesons, we employ the LPSA procedure~\cite{Han:2007ae}, which assumes approximately uniform rapidities within the kinematically allowed interval. We draw independent $\xi_i\in[0,1]$ and define
\begin{equation}
Y_i = Z + \xi_i\,Y \qquad (i=1,\dots,N),
\end{equation}
where the auxiliary parameters $Z$ and $Y$ are fixed by energy-momentum conservation in the hadronization rest frame (\textit{i.e.} $q_k \bar{q}_k$ rest frame):
\begin{equation}
\sum_{i=1}^{N} E_i = \sqrt{s},\qquad \sum_{i=1}^{N} p_{L,i}=0\,.
\end{equation}

The transverse momenta are modeled with a Gaussian distribution subject to momentum conservation:
\begin{equation}
f(\vec p_{T,1},\dots,\vec p_{T,N})\propto
\left[\prod_{i=1}^{N}\exp\!\left(-\frac{\vec p_{T,i}^{\,2}}{\bar\sigma^{\,2}}\right)\right]\,
\delta^{(2)}\!\left(\sum_{i=1}^{N}\vec p_{T,i}\right).
\label{eq:pTdist}
\end{equation}
The width parameter $\bar\sigma$ determines the characteristic transverse spread. In the SM sector, the reference width $\bar{\sigma}_{\text{SM}} = 0.36~\text{GeV}$~\cite{Sjostrand:2006za}.
For the DM case, we assume the transverse kinematics scale linearly with the confinement energy to preserve the event shape topology. Identifying the confinement scale with the dark meson mass, $\Lambda_D = m_{\pi_D}$, we define the scaled width parameter as:
\begin{equation}
    \bar\sigma(\Lambda_D) \equiv \bar\sigma_{\rm SM}\,\frac{\Lambda_{D}}{\Lambda_{\rm QCD}}~,
    \label{eq:sigmabarScale}
\end{equation}
where we adopt $\Lambda_{\rm QCD} = 0.25$ GeV as the conventional hadronic scale~\cite{Sjostrand:2006za}. 
This scaling ensures that the jet-like structure of the DM events mimics QCD behavior but at a different mass scale. In the following, we focus on the resulting single-meson energy distributions, since these are the most directly relevant observables for assessing the impact of the benchmark multiplicity prescription on the downstream flux and event-rate calculations.

\begin{figure}[t]
    \centering
    \includegraphics[width=0.45\textwidth]{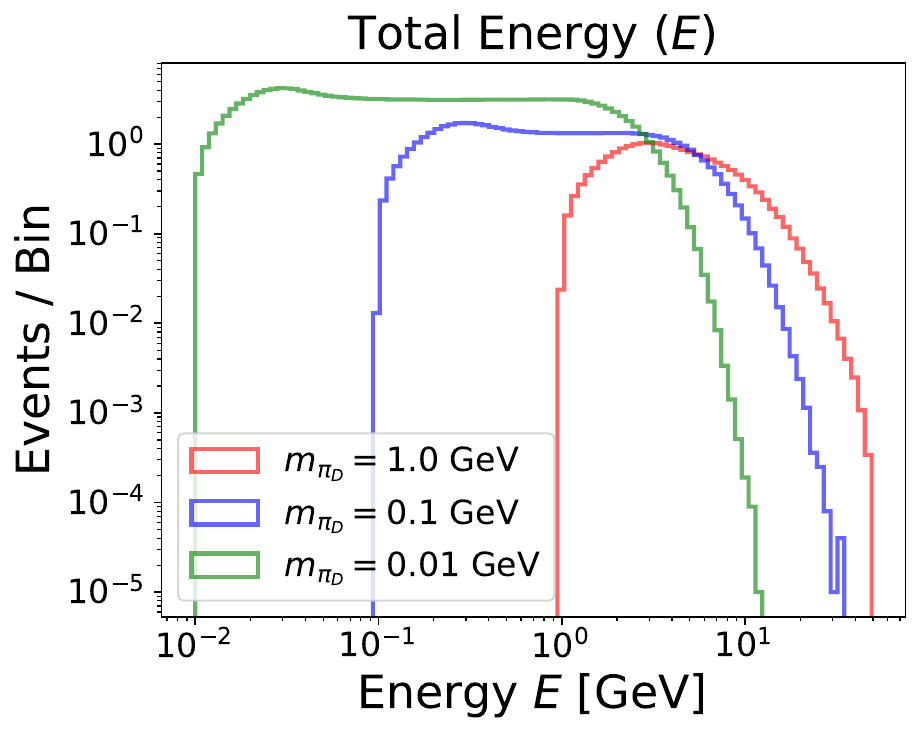} \
    \includegraphics[width=0.45\textwidth]{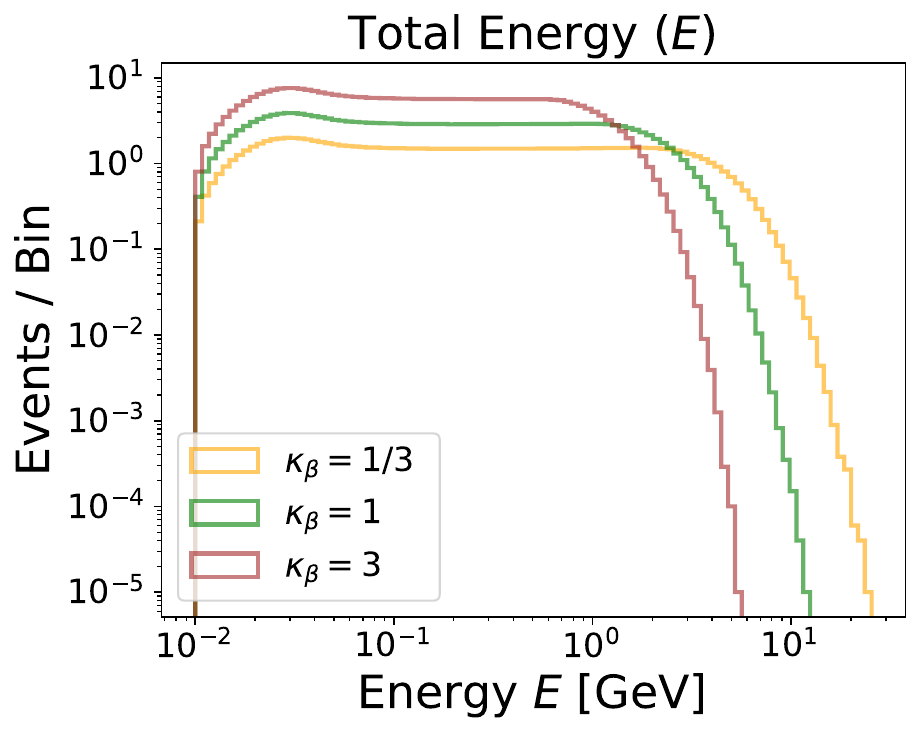} \
    \caption{
    Energy distributions of the produced dark mesons in the $q_k\bar{q}_k$ rest frame for a center-of-mass energy $\sqrt{s}=100$~GeV. Left: dark meson energy spectra for three benchmark dark meson masses, $m_{\pi_D}=1.0$~GeV, $0.1$~GeV, and $0.01$~GeV, evaluated with $\kappa_\beta=1$. Right: the dark meson energy spectrum under the rescaling $\beta(\Lambda_D)\to \kappa_\beta\,\beta(\Lambda_D)$ for a representative benchmark $m_{\pi_D}=0.01$~GeV, with $\kappa_\beta=1/3,\,1,$ and $3$.}

    \label{fig:hv_kinematics}
\end{figure}

The constructed meson energy spectra at fixed $\sqrt{s}=100$~GeV are shown in Figure~\ref{fig:hv_kinematics}. In the left panel, the energy spectra for three benchmark dark meson masses illustrate the expected shift toward lower single-meson energies for lighter dark sectors, reflecting the larger average multiplicity generated by the MQCM prescription. In the right panel, we examine the robustness of this energy-spectrum prediction under the order-one rescaling $\beta(\Lambda_D)\to \kappa_\beta\,\beta(\Lambda_D)$ for a representative benchmark $m_{\pi_D}=0.01$~GeV. As $\kappa_\beta$ increases, the average meson multiplicity becomes larger and the single-meson spectrum shifts smoothly toward lower energies, as expected from energy sharing among a larger number of final-state mesons. This robustness check is motivated by the fact that, for a fixed partonic energy, a change in the average hadron multiplicity modifies the energy sharing among final-state dark mesons.

With the momenta generated in the center-of-mass frame, the kinematics of all dark mesons are fully determined. Each of the meson transverse mass is given by $m_{T,i}=\sqrt{m_{\pi_D}^2+\vec p_{T,i}^{\,2}}$, and the four-momentum in the hadronization rest frame is defined as $p_i^{\mu,\text{rest}}=(E_i,\ \vec p_{T,i},\ p_{L,i})$. To interface with the laboratory frame production mechanism, we perform a Lorentz boost on the entire final state: $p_i^{\mu,\text{lab}} = \Lambda^\mu{}_\nu(\beta_{Z'})\,p_i^{\nu,\text{rest}}$, where the boost vector $\vec{\beta}_{Z'}$ is determined by the (off-shell) $Z'$ boson's four-momentum. The resulting set $\{p_i^{\mu,\text{lab}}\}$ constitutes the dark-matter beam used for the subsequent flux convolution.
\section{Productions of Dark Gauge Boson and Dark Meson} \label{sec:4}

A comprehensive analysis of DM production channels in the upper atmosphere is central to the proposed search strategy. In this section, we detail the cross-section calculations for the three primary production mechanisms: proton bremsstrahlung, the decay of SM mesons, and the Drell-Yan process.

\subsection{Proton Bremsstrahlung Production}
\label{subsec:proton_brem}

We consider the production of a light vector boson $Z'$ in cosmic-ray collisions via the bremsstrahlung-like process
$p + N \to p + N + Z'$, where $N$ denotes an atmospheric nucleon. For simplicity, we use the (total) $pp$ scattering
cross section $\sigma_{pp}$ as an effective hadronic input to obtain an order-of-magnitude estimate.

In the laboratory frame, the incoming proton has energy $E_p$ and three-momentum magnitude
\begin{equation}
P \equiv |\vec p_p| = \sqrt{E_p^2-m_p^2}\, .
\end{equation}
The emitted $Z'$ is characterized by its laboratory energy $E_{Z'}$ and transverse momentum $p_T$. Defining
\begin{equation}
p_{Z'} \equiv |\vec p_{Z'}| = \sqrt{E_{Z'}^2-m_{Z'}^2}\, ,\qquad
p_L = \sqrt{E_{Z'}^2-m_{Z'}^2-p_T^2}\, ,
\end{equation}
we introduce the longitudinal momentum fraction
\begin{equation}
z \equiv \frac{p_L}{P}
= \frac{\sqrt{E_{Z'}^2-m_{Z'}^2-p_T^2}}{\sqrt{E_p^2-m_p^2}} \, ,
\label{eq:z_def}
\end{equation}
and the kinematic combination
\begin{equation}
H \equiv p_T^2 + (1-z)m_{Z'}^2 + z^2 m_p^2 \, .
\label{eq:H_def}
\end{equation}

Following Ref.~\cite{Asai:2022zxw}, we write the double-differential production cross section in variables $(z,p_T^2)$ as
\begin{equation}
\frac{d^2\sigma_{\rm brem}}{dz\, dp_T^2}
=
\sigma_{pp}(s')\;
w_{ba}(z,p_T^2)\;
\left|F_1(m_{Z'}^2)\right|^2\,,
\label{eq:diff_rate_z}
\end{equation}
where $F_1$ is an effective nucleon electromagnetic form factor modeled using the vector-meson dominance prescription. The reduced invariant mass squared after radiating energy $E_{Z'}$ is taken to be
\begin{equation}
s' \simeq 2m_p\left(E_p-E_{Z'}+m_p\right)\,,
\end{equation}
while the initial fixed-target invariant mass squared is
\begin{equation}
s = 2m_p\left(E_p+m_p\right)\,.
\end{equation}

To obtain the spectrum in $E_{Z'}$, we transform from $z$ to $E_{Z'}$ at fixed $p_T^2$ using Eq.~\eqref{eq:z_def}.
The corresponding Jacobian is
\begin{equation}
\left|\frac{\partial z}{\partial E_{Z'}}\right|
=
\frac{E_{Z'}}{z\,(E_p^2-m_p^2)}\,,
\label{eq:jacobian}
\end{equation}
so that
\begin{equation}
\frac{d^2\sigma_{\rm brem}}{dE_{Z'}\, dp_T^2}
=
\sigma_{pp}(s')\;
w_{ba}(z,p_T^2)\;
\frac{E_{Z'}}{z\,(E_p^2-m_p^2)}\;
\left|F_1(m_{Z'}^2)\right|^2\,.
\label{eq:diff_rate_Ezp}
\end{equation}

We then define the \emph{multiplicity per $pp$ collision} as
\begin{equation}
\frac{d^2N_{\rm brem}}{dE_{Z'}\, dp_T^2}
\equiv
\frac{1}{\sigma_{pp}(s)}\;
\frac{d^2\sigma_{\rm brem}}{dE_{Z'}\, dp_T^2}
=
\frac{\sigma_{pp}(s')}{\sigma_{pp}(s)}\;
w_{ba}(z,p_T^2)\;
\frac{E_{Z'}}{z\,(E_p^2-m_p^2)}\;
\left|F_1(m_{Z'}^2)\right|^2 ~,
\label{eq:dNdEdpT2}
\end{equation}
where the splitting kernel for emission of a massive vector boson is
\begin{align}
w_{ba}(z,p_T^2) = \frac{g_{\rm SM}^2}{16\pi^2 H} \Bigg[
& 2 \frac{1+(1-z)^2}{z} 
- 4z(1-z) \frac{2m_p^2 + m_{Z'}^2}{H} \nonumber \\
& + \frac{4z(1-z)}{H^2} \bigg\{ z^2 m_p^4 + (1-z)m_{Z'}^4 + (z^2 - 2z + 2) m_p^2 m_{Z'}^2 \bigg\}
\Bigg]\,.
\label{eq:wba_kernel_simplified}
\end{align}

To remain within the validity regime of the relativistic approximation used in Ref.~\cite{Asai:2022zxw}, we impose the
fixed numerical cuts $E_p>3m_p$, $E_{Z'}>3m_{Z'}$, and $p_T<(E_p-E_{Z'})/3$. In addition, we apply the virtuality cut
used there. The minimum virtuality of the exchanged photon is estimated as
\begin{equation}
|q^2_{\min}|
\simeq
\frac{z^2(1-z)^2}{4E_p^2}\,H^2\, .
\label{eq:qmin}
\end{equation}
We then require $|q^2_{\min}|<\Lambda_{\rm QCD}^2$ with $\Lambda_{\rm QCD}=0.25~\mathrm{GeV}$.

\begin{figure}[htbp]
    \centering
    \includegraphics[width=0.48\textwidth]{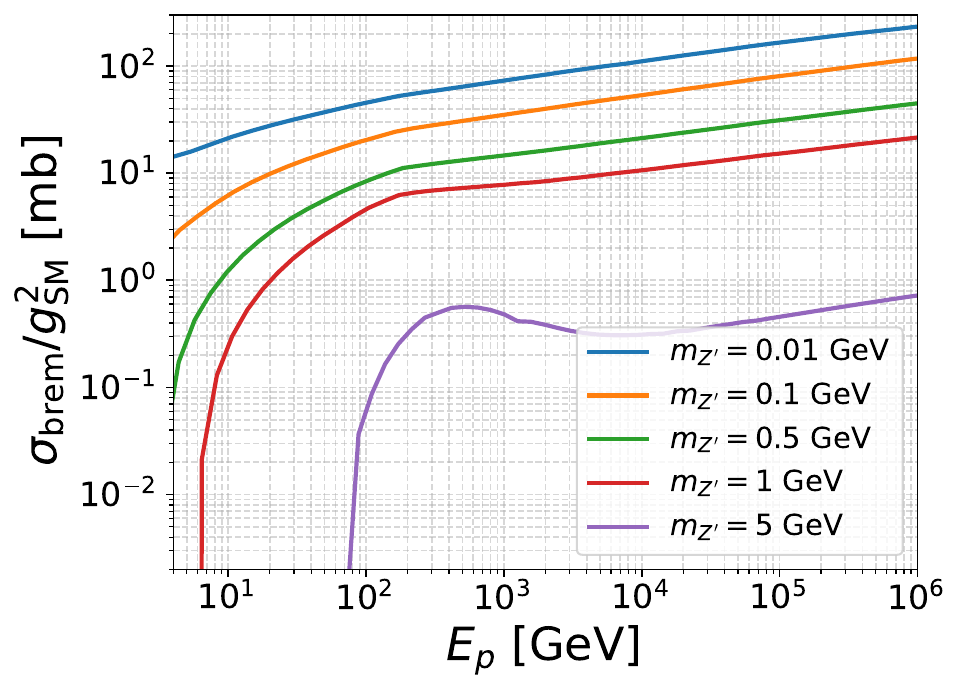}
    \hfill
    \includegraphics[width=0.48\textwidth]{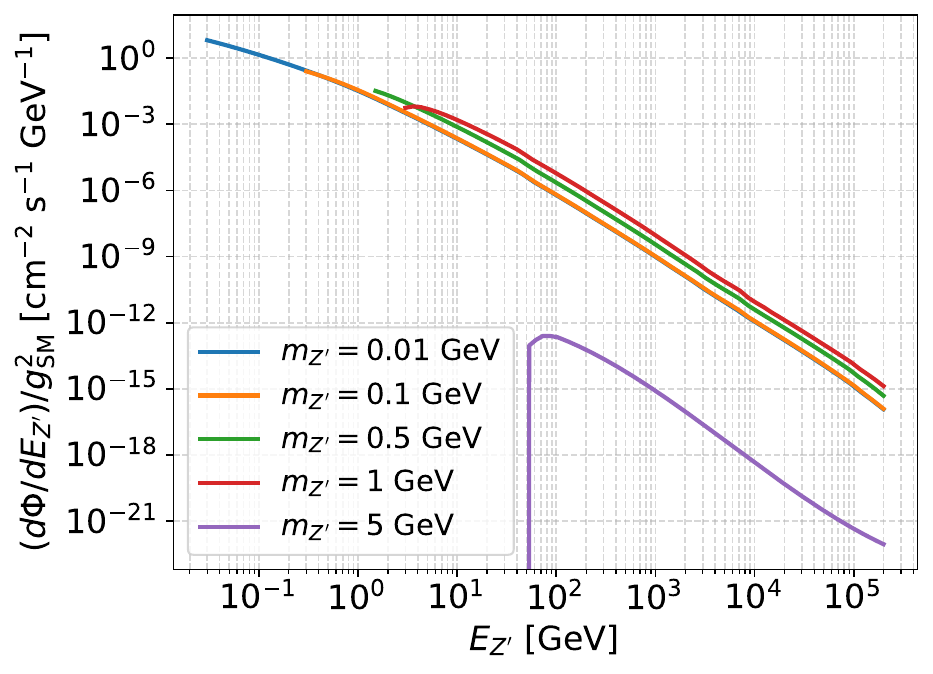}
    \caption{
        \textbf{Left:} The bremsstrahlung production cross-section $\sigma(pp \to p p Z')$ as a function of the incident proton energy
        $E_p$ for $m_{Z'} = 0.01,\,0.1,\,0.5,\,1.0,\,5.0$~GeV, computed with the benchmark coupling $g_{SM}=1$
        (to be rescaled when deriving limits).
        \textbf{Right:} The corresponding differential flux of $Z'$ at the production point as a function of $E_{Z'}$,
        obtained from the multiplicity convolution in Eq.~\eqref{eq:Phi_Zp}.
    }
    \label{fig:cross_section_flux}
\end{figure}

The total bremsstrahlung production cross section $\sigma(pp\to ppZ')$ is obtained by integrating
Eq.~\eqref{eq:diff_rate_Ezp} over the allowed phase space, and is shown in the left panel of Fig.~\ref{fig:cross_section_flux}.
Finally, we convolve the per-collision multiplicity with the isotropic cosmic-ray proton flux $\phi_p(E_p)$ to obtain the
differential $Z'$ flux at the production point,
\begin{equation}
\Phi_{Z'}(E_{Z'})
=
2\pi \int_{E_{p,\min}}^{\infty} dE_p\;\phi_p(E_p)\;
\int dp_T^2\;
\frac{d^2N_{\rm brem}}{dE_{Z'}\, dp_T^2}\,,
\label{eq:Phi_Zp}
\end{equation}
where the prefactor $2\pi$ corresponds to integrating an isotropic differential flux over the downward hemisphere.
The resulting spectra are presented in the right panel of Fig.~\ref{fig:cross_section_flux}.

\subsection{SM Meson Decay}
Cosmic-ray proton interactions with atmospheric nuclei initiate hadronic cascades at energies exceeding a few GeV, producing copious amounts of pseudoscalar mesons, primarily $\pi^0$ and $\eta$. In scenarios involving a light dark sector, these SM mesons serve as a potent source of dark gauge bosons ($Z^\prime$) via rare radiative decays, provided $m_{Z^\prime} < m_{\pi,\eta}$. We focus on the mass hierarchy where the dark gauge boson is more than twice as heavy as the dark quarks ($m_{Z^\prime} > 2 m_{q_k}$), such that the on-shell decay ($Z^\prime \to q_k \bar{q}_k$) dominates the production of dark sector particles.

The primary production channels for dark gauge bosons from pseudoscalar mesons are the radiative decays  $\pi^0 \to Z^\prime \gamma$ and $\eta \to Z^\prime \gamma$. The branching fractions for these processes, governed by the gauge coupling to the SM fermions and the phase space suppression, are given by:
\begin{align}
\text{Br}(\pi^0 \to \gamma Z^\prime) &= 2 \epsilon^2 \left(1 - \frac{m_{Z^\prime}^2}{m_{\pi^0}^2}\right)^3 \text{Br}(\pi^0 \to \gamma \gamma), \\
\text{Br}(\eta \to \gamma Z^\prime) &= 2 \epsilon^2 \left(1 - \frac{m_{Z^\prime}^2}{m_{\eta}^2}\right)^3 \text{Br}(\eta \to \gamma \gamma),
\end{align}
where $\epsilon=3 g_{\rm SM}/e$~\footnote{The factor of 3 arises from the triangle anomaly. The amplitude for $\pi^0 \to \gamma \gamma$ is proportional to $(Q^{\rm EM}_u)^2 - (Q^{\rm EM}_d)^2 =1/3$, whereas the amplitude for $\pi^0 \to \gamma Z^\prime$ is proportional to $Q^{\rm EM}_u - Q^{\rm EM}_d =1$, given that the up and down quarks are assumed to have the same coupling under $U(1)_D$. Similarly, the corresponding factor for the octet $\eta$ state is also 3, though it is subject to a correction due to octet-singlet mixing. We will neglect this correction effect in our analysis.} in our setup, and the SM meson branching ratios are $\text{Br}(\pi^0 \to \gamma \gamma) \sim 0.99$ and $\text{Br}(\eta \to \gamma \gamma) \sim 0.39$~\cite{ParticleDataGroup:2024cfk}. We assume that the dark photon subsequently decays into dark quarks with a branching ratio of $\text{Br}(Z^\prime \to q_k \bar{q}_k) \sim 1$, which is valid when $g_{\rm SM}$ is small enough that decays to SM fermions are suppressed relative to the dark sector channel (assuming $g^\prime \gg g_{\rm SM}$).

\begin{figure}[h]
    \centering
    \includegraphics[width=0.45\textwidth]{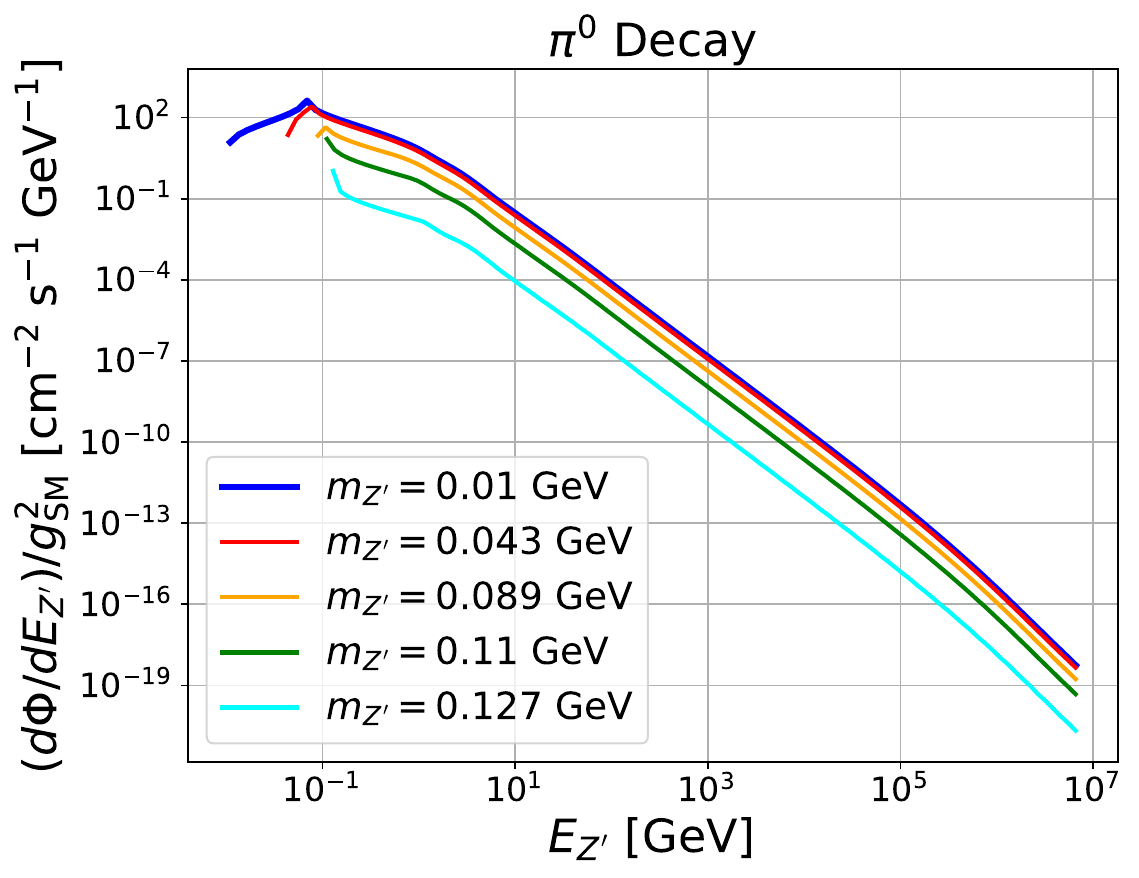}
    \includegraphics[width=0.45\textwidth]{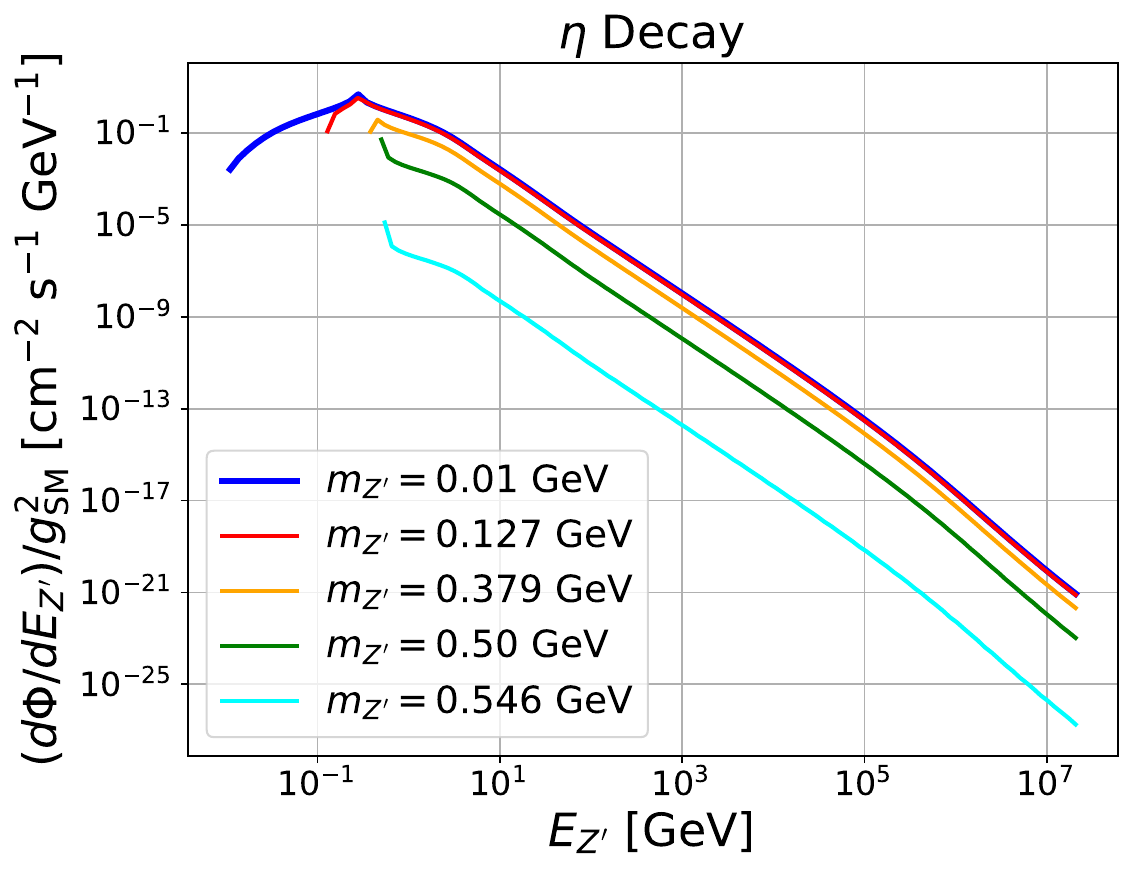}
    \caption{\label{fig:mesondecay} $Z^\prime$ Flux from $\pi^0$ and $\eta$ decay}
\end{figure}

To model the production of atmospheric mesons, we adopt the calculation framework implemented in the \textsc{HeavenlyMCP} code~\cite{ArguellesDelgado:2021lek}. This tool incorporates the Matrix Cascade Equation (MCEq) package~\cite{Fedynitch:2015zma} to solve for the depth-dependent meson fluxes, utilizing the cosmic ray spectrum model from~\cite{Gaisser:2011klf}, the hadronic interaction model~\cite{Fedynitch:2018cbl} and the atmospheric model~\cite{https://doi.org/10.1029/2002JA009430}.
While \textsc{HeavenlyMCP} was originally designed to compute the flux of millicharged particles via the three-body decay of SM mesons, we have adapted the framework to instead simulate the two body radiative decays $\pi^0 \to Z^\prime \gamma$/$\eta \to Z^\prime \gamma$. 
The dark gauge boson fluxes for different mediator masses are presented in Figure~\ref{fig:mesondecay}. 

Finally, the dark gauge boson flux is converted into a flux of dark pions via the hadronization process described in Section~\ref{sec:3}. Since the dark quarks are produced from the decay of an on-shell $Z^\prime$, the center-of-mass energy for the hadronization process in the Modified Quark Combination Model is fixed at $\sqrt{s} = m_{Z^\prime}$. 
The dark pion fluxes at the Earth’s surface resulting from $\pi^0$ and $\eta$ decays are presented below in this section.

\subsection{Drell-Yan Production of Dark Quark}
\label{subsec:Drell-Yan}
Dark quarks can be produced via the $s$-channel exchange of a virtual $Z^\prime$ through quark-antiquark parton scattering, $q\bar{q} \to Z^{\prime *} \to q_k \bar{q}_k$. This Drell-Yan production process is particularly significant when the mass of the dark gauge boson is $\gtrsim \mathcal{O}(1)$ GeV.
We calculate the corresponding differential cross sections for dark quark pair production using \texttt{MadGraph5}~\cite{Alwall:2014hca}, incorporating UFO model files generated by \texttt{FeynRules}~\cite{Alloul:2013bka}. 
In our simulation, the dark sector gauge coupling is set to unity, while the SM coupling $g_{\rm SM}$ is fixed at $10^{-2}$.
This configuration reflects the parameter space of interest, characterized by $g^\prime \sim 1$ and $g_{\rm SM} \ll 1$, where the width of the $Z^\prime$ is dominated by the decay channel $Z^\prime \to q_k \bar{q}_k$.
Reducing  $g_{SM}$ to a smaller value does not affect the kinematic distributions presented in this work, as it serves primarily as a normalization factor in this regime.  

\begin{figure}[h]
    \centering
    \includegraphics[width=0.3\textwidth]{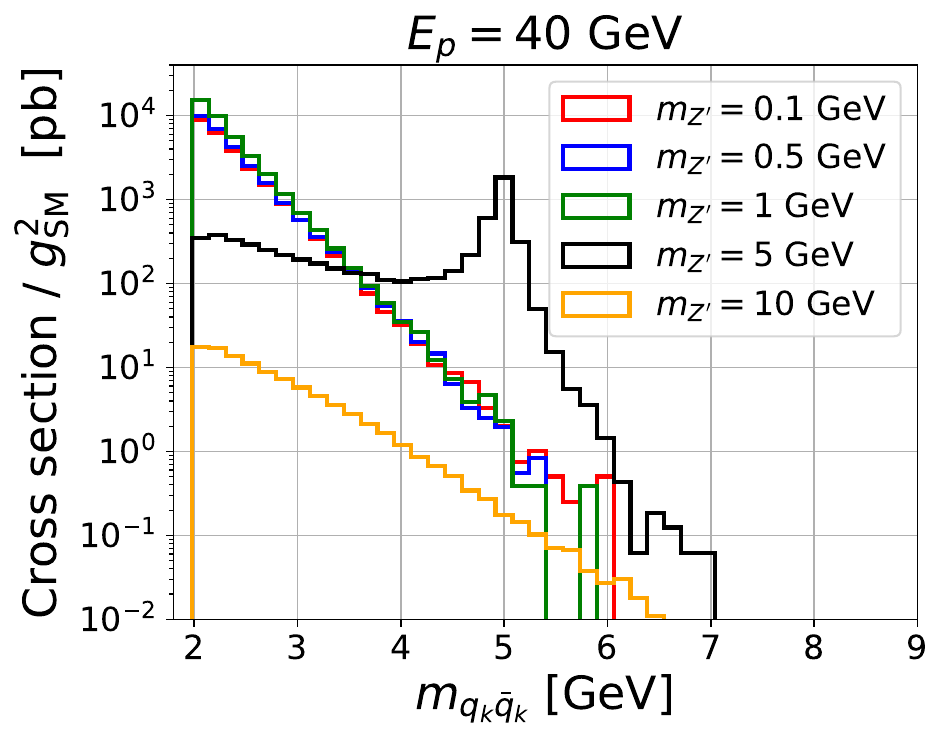}
    \includegraphics[width=0.3\textwidth]{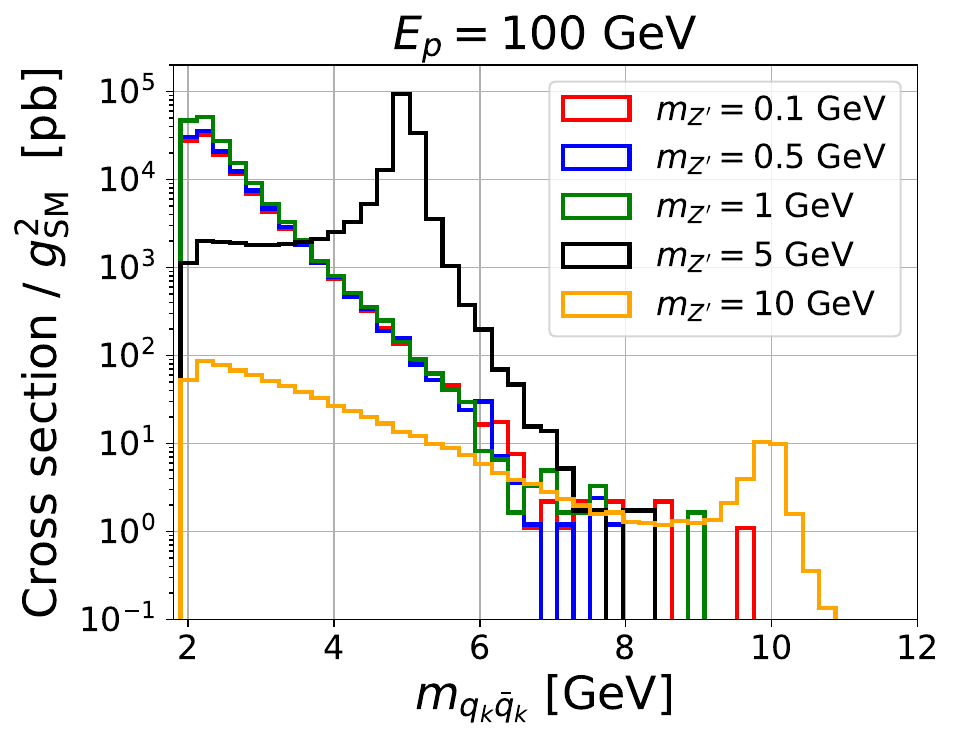}
    \includegraphics[width=0.3\textwidth]{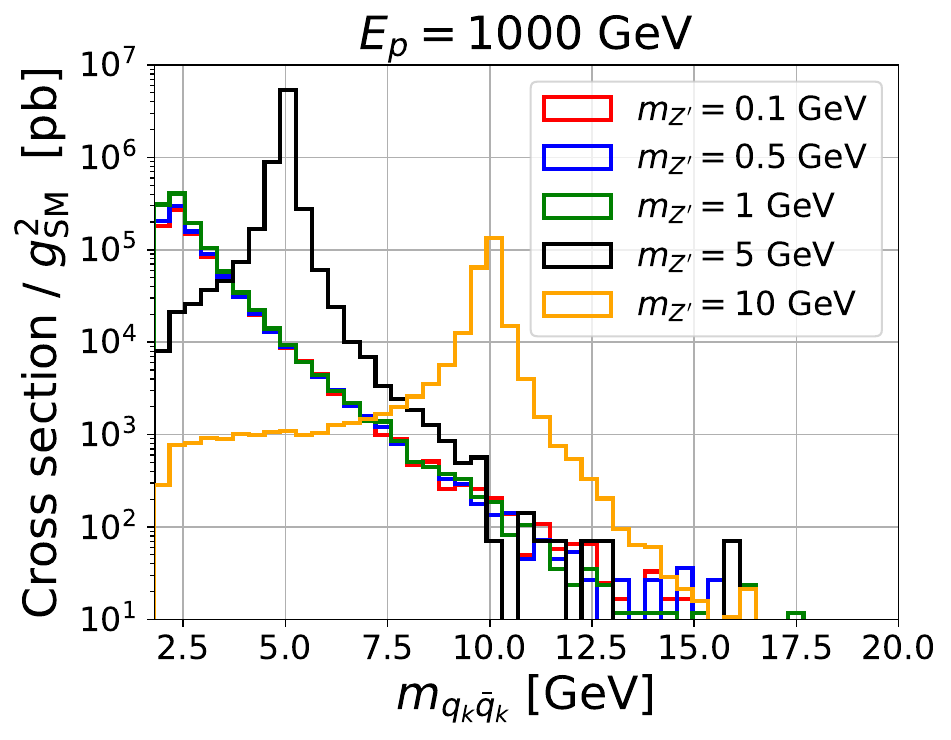}
    \caption{\label{fig:mxxE} The differential production cross section of the Drell-Yan process for different incident proton energies ($E_p$). From left to right, $E_p = 40, 100,$ and $1000$ GeV. }
\end{figure}

All events are simulated in the Laboratory frame, assuming a fixed-target configuration with one of the protons at rest. In Figure~\ref{fig:mxxE}, we present the invariant mass distribution of the dark quark pair for various incident proton energies.  
It is observed that the differential cross section exhibits a weak dependence on the $Z^\prime$ mass in the regime where $m_{Z^\prime} \lesssim \mathcal{O}(1)$ GeV.  
However, a significant resonant enhancement appears when the center-of-mass energy of the colliding protons, $\sqrt{s} \sim \sqrt{2 m_p E_p}$, exceeds the $Z^\prime$ mass. This invariant mass peak becomes increasingly pronounced at higher collision energies, attributed to the reduced suppression from Parton Distribution Functions (PDFs) at the relevant momentum fractions. 
Specifically, for the lower incident energy of $E_p = 40$ GeV, the production of heavier mediators (e.g., $m_{Z^\prime} = 5, 10$ GeV) is kinematically forbidden or highly suppressed. As $E_p$ increases to 1000 GeV, the resonance peaks for all considered masses become clearly distinguishable. Furthermore, in the low invariant mass region ($m_{q_k \bar{q}_k} \ll m_{Z^\prime}$), the cross section exhibits a continuum behavior dominated by off-shell exchange, which explains the overlapping curves for light $Z^\prime$ scenarios.

\begin{figure}[h]
    \centering
    \includegraphics[width=0.45\textwidth]{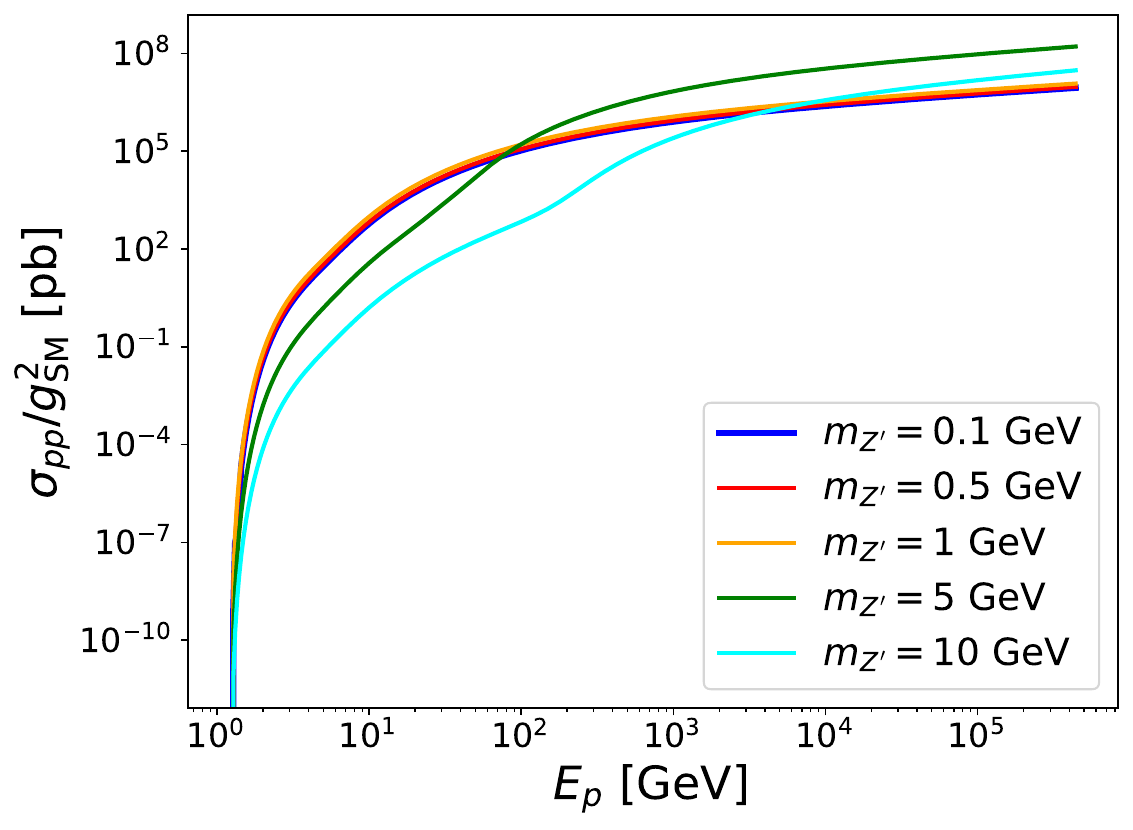}
    \includegraphics[width=0.45\textwidth]{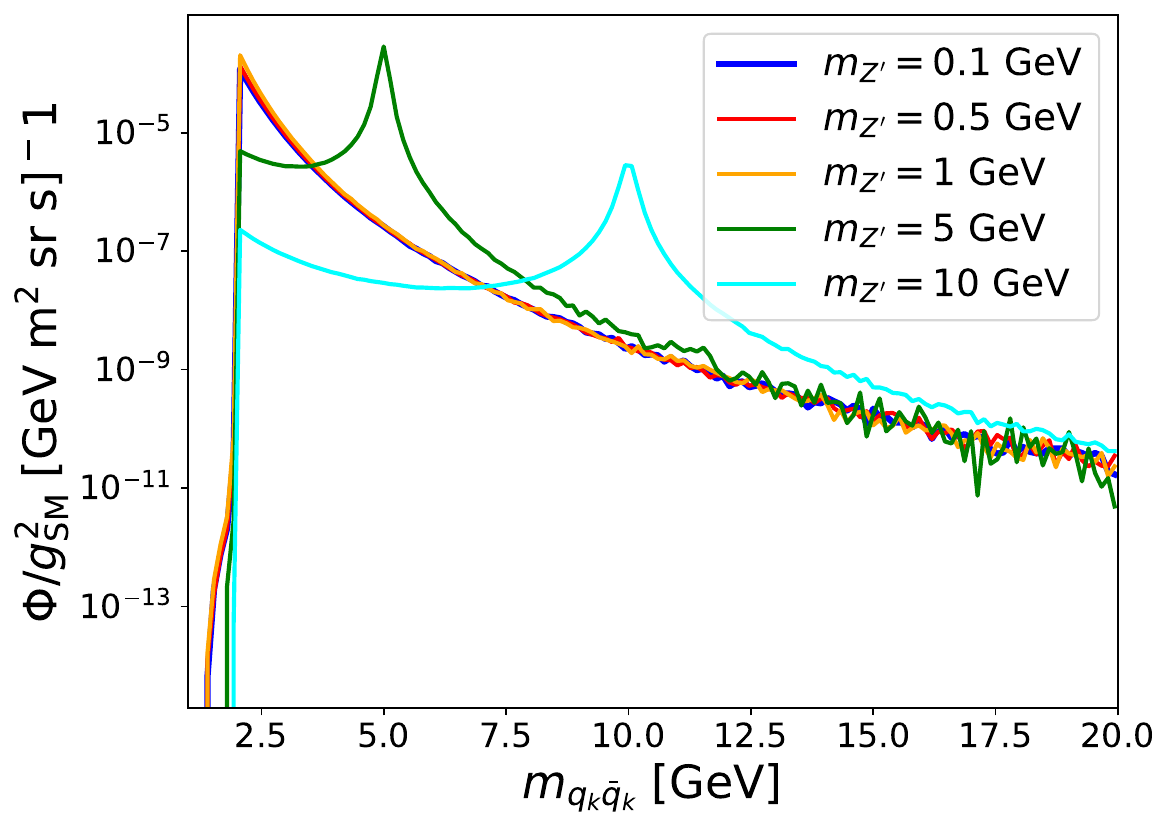}
    \caption{\label{fig:xsecDY} Left: Total production cross section of the Drell-Yan process as a function of incident proton energy. Right: The differential flux in terms of the dark quark pair invariant mass $m_{q_k \bar{q}_k}$ due to the Drell-Yan process. }
\end{figure}

In the left panel of Figure~\ref{fig:xsecDY}, we illustrate the total cross section for the Drell-Yan process, normalized to unit couplings $g_{\rm SM}=1$. The dark matter gauge coupling $g^\prime=1$, and the mass of the $Z^\prime$ boson is indicated in the legend. While the dark quark mass is set to $m_{Z^\prime} /100$ in this simulation, we found that the total cross section is insensitive to the specific dark quark mass, provided it remains significantly lighter than the $Z^\prime$. 
As indicated in the figure and consistent with the differential distributions in Figure~\ref{fig:mxxE}, the total cross sections are nearly identical for $m_{Z^\prime} \lesssim \mathcal{O}(1)$ GeV. 
For heavier $Z^\prime$ masses, the cross section drops sharply at lower proton energies where the center-of-mass energy is insufficient to produce the on-shell resonance. As the incident energy increases, the cross sections for different masses tend to converge and saturate. In this energy regime, the production rate is mainly governed by the couplings and the overall PDF scaling. 

The differential flux from the Drell-Yan process is then straightforwardly obtained by convoluting the production cross section with the cosmic ray proton flux $\phi_p(E_p)$:
\begin{align}
\Phi_{Z^\prime} (m_{q_k \bar{q}_k})=\int dE_p \phi_p(E_p) \frac{1}{\sigma^{\rm tot}_{pp} (E_p)} \frac{d\sigma^{\rm DY}_{pp}(E_p)}{dm_{q_k \bar{q}_k}}~,~
\end{align}
where $\sigma^{\rm tot}_{pp} (E_p)$ is the total proton-proton collision cross section, taken from the Particle Data Group~\cite{ParticleDataGroup:2024cfk}.  
The corresponding fluxes for different $Z^\prime$ masses are plotted in the right panel of Figure~\ref{fig:xsecDY}. 
Similar to the cross sections, the flux dependence on $m_{Z^\prime}$ is quite weak for $m_{Z^\prime} \lesssim \mathcal{O}(1)$ GeV. 
For heavier $Z^\prime$, although the flux of high-energy cosmic ray protons is steeply falling, the resonance peak remains prominent. In fact, even after accounting for the power-law suppression of the cosmic ray spectrum, the dark quark pair flux is still dominated by production at the $Z^\prime$ mass pole for $m_{Z^\prime}=5$ and $10$ GeV.

\subsection{Overall Flux of Dark Meson}
\label{sec:overall_flux_dark_meson}

Sections~\ref{subsec:proton_brem}-\ref{subsec:Drell-Yan} provide the differential partonic flux from the relevant atmospheric production mechanisms considered in this work. For a fixed mediator mass $m_{Z'}$, we construct the total partonic flux by summing the channel contributions at the level of the same laboratory-frame observable (here taken to be the $Z'$ total energy $E_{Z'}$). This total $Z'$ flux~\footnote{An off-shell $Z^{\prime *}$ is referred to for the Drell-Yan production channel.} is the only production input needed for calculating the dark meson flux. Note that we assume the dark matter coupling $g^\prime \gg g_{\rm SM}$, so that the $Z^\prime$ dominantly decays into the dark quark pair.  

Given $(m_{q_k \bar{q}_k},m_{\rm \pi_D})$, we then apply the event-level prescription of Section~\ref{sec:3} to map each produced partonic configuration into a set of final-state dark-meson four-momenta, and thereby obtain the dark-meson energy spectrum. 

Assuming three nearly degenerate dark quark flavors, the dark hadronization process yields a total of nine dark meson states, including the SU(3) octet and a singlet. {Although the dark $\eta_D^\prime$ mesons might be heavier analogously to the SM}, the exact mass spectrum and fragmentation dynamics in the dark sector are highly model-dependent. As a simple and conservative estimate, we assume that all nine meson states are produced with similar probabilities and kinematics, and that they are sufficiently long-lived to be stable on the length scale of the experiment (several tens of kilometers). However, based on our $U(1)_D$ charge assignment introduced in Section~\ref{sec:meson}, only four dark kaons ($K_D$) out of these nine states couple to the SM particles through the $Z^\prime$ portal. Consequently, a factor of 4/9 is explicitly imposed when evaluating the interacting stable meson flux.
In what follows, we use $E_{K_D}$ to denote the total energy of a single dark meson $K_D$ in the laboratory frame. 
The resulting observable is the differential flux $d\Phi/dE_{K_D}$, reported throughout normalized by $g_{\text{SM}}^2$ so that alternative coupling choices can be obtained by a simple rescaling.

\begin{figure}[h]
    \centering
    \includegraphics[width=0.45\textwidth]{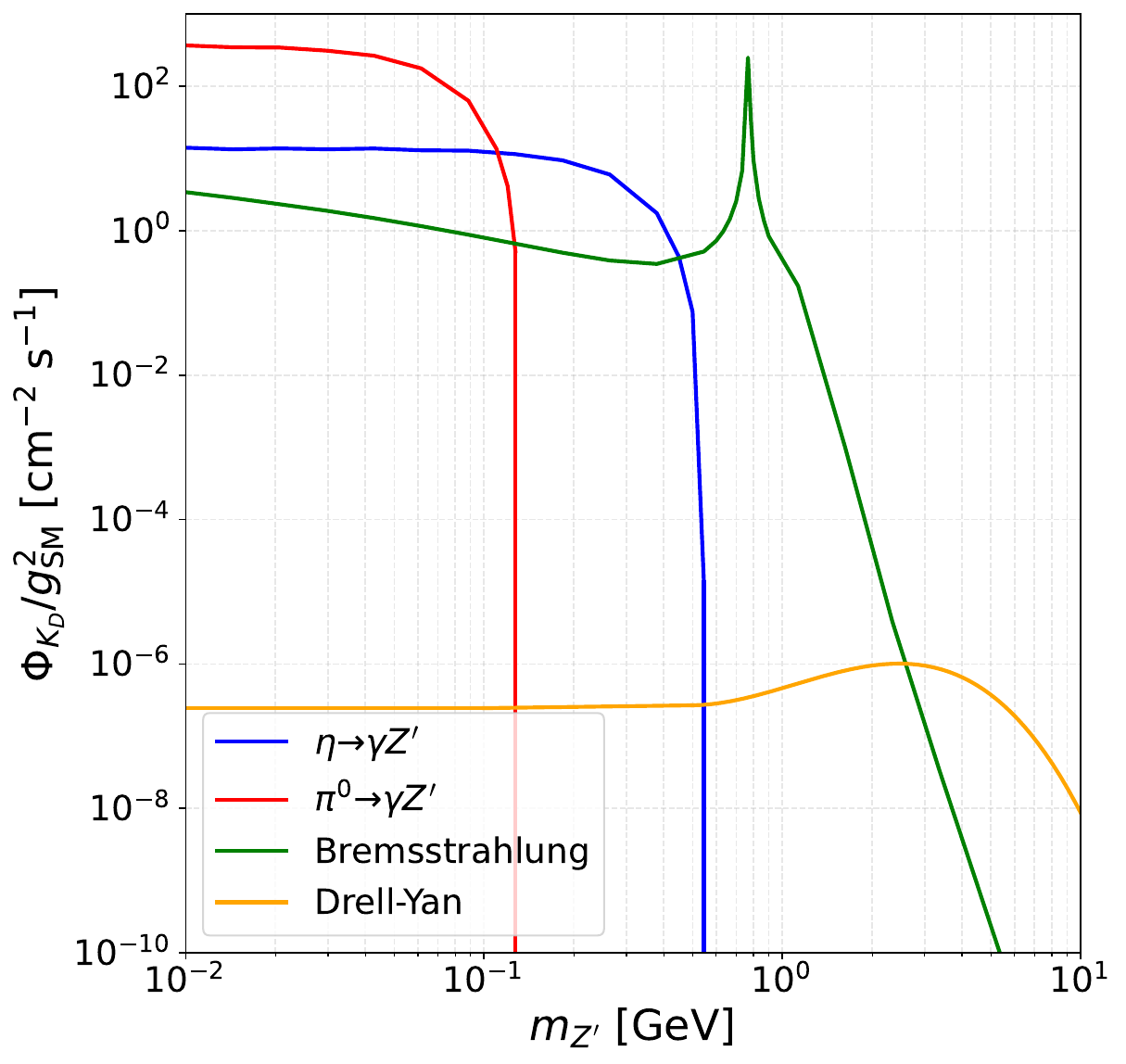}\
    \includegraphics[width=0.45\textwidth]{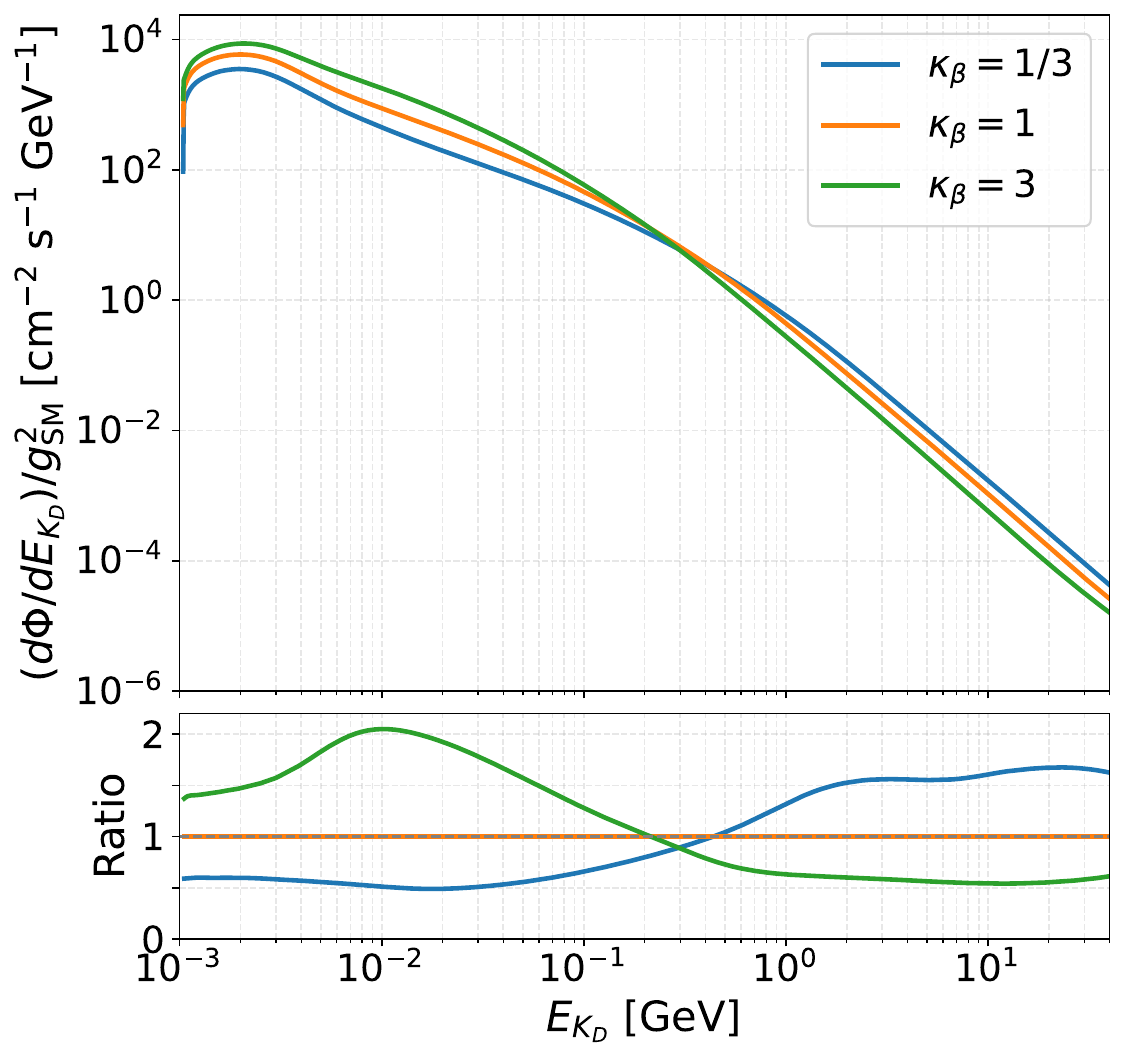}
    \caption{\label{fig:integrate_flux}
    Left: energy-integrated dark-meson flux (integrated over $E_{K_D}$) as a function of the mediator mass $m_{Z'}$, shown decomposed into the contributions associated with the individual production mechanisms and normalized by $g_{\rm SM}^2$.
   Right: robustness of the total dark-meson differential flux $(d\Phi/dE_{K_D})/g_{\rm SM}^2$ under the  rescaling $\beta(\Lambda_D)\to \kappa_\beta\,\beta(\Lambda_D)$ for $m_{Z'}=0.1~\mathrm{GeV}$. The upper subpanel shows the total flux for $\kappa_\beta=1/3,\,1,$ and $3$, while the lower subpanel displays the ratios to the fiducial case $\kappa_\beta=1$. }
\end{figure}

Figure~\ref{fig:integrate_flux} summarizes the final output of this pipeline. The left panel shows the energy-integrated dark-Kaon ($K_D$) flux as a function of $m_{Z'}$, where the integration is performed over $E_{K_D}$ for the benchmark relation $m_{Z^\prime}= 100  m_{q_k}$ and $m_{\pi_D}=2 m_{q_k}$. The decomposition highlights how the relative importance of the individual $Z'$ production mechanisms propagates into the final $K_D$ yield once the hadronization mapping is applied. In particular, the meson-decay contributions exhibit the expected kinematic shutoff once $m_{Z'}$ exceeds the parent-meson mass, while the remaining channels continue to contribute at larger $m_{Z'}$.

The right panel of Figure~\ref{fig:integrate_flux} shows the total differential flux $(d\Phi/dE_{K_D})/g_{\rm SM}^2$ for $\kappa_\beta=1/3,\,1,$ and $3$, together with the corresponding ratios to the fiducial case $\kappa_\beta=1$, for a representative benchmark $m_{Z'}=0.1~\mathrm{GeV}$. The overall spectral falloff follows from the steeply decreasing cosmic-ray spectrum encoded in the parent $Z'$ flux, while the $\kappa_\beta$ scan shows that the benchmark multiplicity prescription affects both the normalization and the shape of the dark-meson spectrum. Increasing $\kappa_\beta$ enhances the low-energy part of the flux and suppresses the high-energy tail, as expected from energy sharing among a larger number of final-state dark mesons. Unless otherwise stated, the detector-level scattering calculations in the next section are performed with the fiducial choice $\kappa_\beta=1$, while the $\kappa_\beta$ scan shown here serves to illustrate the associated theoretical uncertainty.

Before using this flux as the detector input, one may in principle account for additional dark meson propagation through the atmosphere and the local rock overburden above JUNO. Recent studies of cosmic-ray-boosted dark matter have pointed out that such attenuation effects can become relevant when the scattering rate in terrestrial matter is sufficiently large~\cite{Xia:2021vbz,Alvey:2022pad}. In the present work, however, our goal is only to determine whether this effect can substantially modify the sensitivity in the small-coupling region of interest, rather than to perform a full transport simulation. We therefore adopt a simplified overburden estimate based on the optical depth,
\begin{equation}
\tau(E)=N_A X \sum_i \frac{w_i}{A_i}\,\sigma_i^{\rm tot}(E),
\end{equation}
with the corresponding survival probability estimated as $P_{\rm surv}(E)=e^{-\tau(E)}$. Here $N_A$ is  Avogadro's number, $X$ denotes the rock column depth, while $w_i$, $A_i$, and $\sigma_i^{\rm tot}$ are the mass fraction, atomic mass, and total dark-meson scattering cross section of the relevant element $i$, respectively. For the local overburden we approximate the rock composition by O/Si-dominated material, with mass fractions $w_{\rm O}\simeq 0.53$ and $w_{\rm Si}\simeq 0.47$, and take a vertical column depth $X\simeq 1.81\times 10^{5}~\mathrm{g/cm^{2}}$ corresponding to a rock overburden of $\sim\!700~\mathrm{m}$ above JUNO. For each element $i$, the total dark-meson scattering cross section is taken as $\sigma_i^{\rm tot}=\sigma_i^{\rm EL}+\sigma_i^{\rm DIS}$. To account for the range of incidence directions, we replace the strictly vertical survival probability by its hemispherical angular average,
\begin{equation}
\bar{P}_{\rm surv}(E) \;=\; 2\int_{0}^{1} \mu\, e^{-\tau_{0}(E)/\mu}\, d\mu\,,
\end{equation}
where $\mu=\cos\theta$ and $\tau_{0}(E)$ is the vertical optical depth given by Eq.~(4.17). The relative flux loss is then defined as
\begin{equation}
\Delta\Phi/\Phi \;\equiv\; 1 - \frac{\int dE\,\Phi(E)\,\bar{P}_{\rm surv}(E)}{\int dE\,\Phi(E)}\,.
\end{equation}
For the representative benchmark points $m_{Z'}=0.1,\,0.01,$ and $0.001~\mathrm{GeV}$ and a reference coupling $g_{\rm SM}=10^{-3}$, the resulting flux losses are at most of order $10^{-8}$, while even at the larger reference coupling $g_{\rm SM}=10^{-2}$ the attenuation remains below the per-mille level for all three benchmark masses. Since our projected JUNO sensitivity lies in the regime $g_{\rm SM}\lesssim 10^{-4}$, where the optical depth is further suppressed, we conclude that the propagation loss in the local overburden is negligibly small and does not affect the projected JUNO reach.

\section{DM-Nucleus Scattering at the JUNO Detector} \label{sec:5} 

As discussed in Section~\ref{sec:2}, the 4/9 of the dark meson species, denoted by $K_D$, interacts with the SM via a vector mediator $Z^\prime$. In our simulation, we adopt the scalar DM tune within GENIE, where the DM particle is a complex scalar coupled to a vector mediator. The interaction Lagrangian describing the coupling of the mediator to the DM and SM fermions in accordance with the definition in GENIE is given by:
\begin{equation}
\mathcal{L}_{\text{int}} = 2\times i g^\prime Z^{\prime \mu} (K_D^* \partial_\mu K_D - \partial_\mu K_D^* K_D) + g_{\text{SM}} Z^\prime_\mu \sum_{f} \bar{\psi}_f \gamma^\mu (Q^f_L P_L + Q^f_R P_R) \psi_f ~,
\end{equation}
where $f/K_D$ denotes the SM fermions / Kaon-like dark mesons, $P_{L,R}$ are the chiral projection operators, and the factor of 2 entering the $K_D$ coupling originates from the charge assignment defined in Eq.~\ref{eq:zkk}. We assume vector-like couplings to the SM fermions such that $Q^f_L = Q^f_R = Q^f_V$. 

\subsection{Simulation Framework with GENIE}
Hadronic interactions of DM share significant kinematic and dynamical similarities with neutral-current (NC) neutrino scattering. Consequently, it is natural to utilize established neutrino Monte Carlo software suites for simulations. We employ the \textbf{GENIE} generator~\cite{Andreopoulos:2009rq, Andreopoulos:2015wxa}, specifically integrating the Boosted Dark Matter (BDM) module~\cite{Berger:2018urf} to perform cross-section calculations and event generation. 

GENIE provides a comprehensive simulation of nuclear effects, including nucleon Fermi motion, Pauli blocking, and Final-State interactions of hadrons as they exit the nuclear remnant. The simulation covers two primary scattering regimes:
\begin{enumerate}
    \item \textbf{Elastic Scattering (EL):} The DM undergoes elastic scattering off nucleons. Under the assumption of a vector coupling, the process is described by the nucleon vector form factors ($F_1$ and $F_2$), which parameterize the charge and magnetic moment distributions. Consistent with the implementation in GENIE, these form factors are assumed to have a dipole form. Their normalizations are constrained by the nucleon electric charges and anomalous magnetic moments. 

    \item \textbf{Deep Inelastic Scattering (DIS):} For high momentum transfer, the interaction enters the DIS regime. In the case of complex scalar dark matter with a vector coupling, the scattering cross-section is described by a hadronic tensor dependent primarily on the $F_1$ and $F_2$ structure functions derived from parton distribution functions (PDFs). To extend validity into the relatively low-$Q^2$ regime characteristic of BDM scenarios, GENIE employs PDFs augmented with Bodek-Yang corrections. The hadronization of the final state is governed by the AGKY model, which selects the fragmentation scheme based on the invariant mass of the hadronic system ($W$). For low $W \lesssim 2.3$ GeV, an empirical KNO-based model tuned to neutrino data is utilized, while for high $W \gtrsim 3$ GeV, the simulation transitions to PYTHIA fragmentation, with a linear transition window in between. 
\end{enumerate}

\subsection{Cross-Section Calculation and Event Generation }
The simulation workflow consists of two primary stages: the pre-computation of cross-section splines and the subsequent generation of scattering events.

First, we generate the scattering cross-section splines using the GENIE \texttt{gmkspl\_dm} application. This step pre-calculates the interaction probabilities for the specific target nucleus. For the JUNO detector, the primary targets are $^{12}\mathrm{C}$ and $^1\mathrm{H}$. 
We employ the \texttt{GDM18\_00b\_00\_000} tune, which defines the physics model for scalar dark matter interactions. 
In this simulation, both the DM and SM couplings are normalized to unity, and the $g_{\rm SM}$ will be rescaled subsequently to derive the sensitivity limits. 
We scan the dark gauge boson ($Z^\prime$) mass across the range of 1 MeV to 10 GeV, with the dark meson mass fixed at 1/50th of the $Z^\prime$ mass. This mass relation is consistent with the configuration used for the flux calculation in the preceding section. 
The splines are generated up to a maximum incident energy of $E_{\text{max}} = 1000$ GeV with 200 knots per spline to ensure high-precision interpolation.

Utilizing the pre-computed cross section splines, event samples are generated via the GENIE \texttt{gevgen\_dm} application. 
To characterize the differential cross-sections across a wide kinematic range, event simulations are performed at discrete incident DM kinetic energies. The energy points are sampled on a logarithmic scale up to 1 TeV. The lower energy thresholds are set to 0.05 GeV for EL and 0.5 GeV for DIS. Flux contributions below these limits were found to yield negligible signal rates, mainly attributed to the suppressed scattering cross section and the signal selection cuts. 

For a specific energy point $E$, the generation procedure is as follows:
\begin{verbatim}
gevgen_dm -n 10000 -m [Dark_Meson_Mass] -g 1 -z 100 -t [Target_PDG] \
  --tune GDM18_00b_00_000 --event-generator-list [Channel] \
  --cross-sections [Spline_File].xml \
  -e $E --seed [Random_Seed] -o dm_[Channel]_E${E}.ghep.root
\end{verbatim}
Here, \texttt{-n} sets the number of events to $10^4$ to ensure sufficient statistics for event counting, and \texttt{-t} specifies the target nucleus code (e.g., \texttt{1000060120} for Carbon-12 in JUNO). The \texttt{--event-generator-list} flag is used to select between \texttt{DMEL} and \texttt{DMDIS}, allowing for the isolation of specific interaction channels. The output is saved in the native GHEP ROOT format and subsequently converted to the \texttt{gxml} summary format using the \texttt{gntpc} utility for analysis.

\subsection{Signal Selection and Sensitivity at JUNO}

The total expected signal yield in the JUNO detector, $N_{\text{total}}$, is obtained by integrating the product of the flux and the scattering cross section over the incoming dark meson energy $E_{K_D}$:
\begin{equation}
N_{\text{total}} = T \times \int_{E_{\text{min}}}^{E_{\text{max}}} \frac{d\Phi}{dE_{K_D}} \times \left[ N_{\text{H}} \cdot \sigma_{\text{H}}^{\text{vis}}(E_{K_D}) + N_{\text{C}} \cdot \sigma_{\text{C}}^{\text{vis}}(E_{K_D}) \right] dE_{K_D}~.~
\label{eq:total_events}
\end{equation}
Here, $T$ denotes the experimental exposure time (assumed to be 1 year), and $d\Phi/dE_{K_D}$ represents the differential flux of the dark meson. 
The terms $N_{\text{H}}$ and $N_{\text{C}}$ correspond to the number of target Hydrogen and Carbon-12 nuclei contained within the 20 kton fiducial mass of JUNO, respectively. 
The quantities $\sigma_{\text{H}}^{\text{vis}}(E_{K_D})$ and $\sigma_{\text{C}}^{\text{vis}}(E_{K_D})$ denote the effective visible scattering cross sections. They incorporate the detector response through the reconstruction of the visible final-state energy and, for recoil protons, the quenching effect described by Birks' law~\cite{vonKrosigk:2013sa}. Since JUNO is a liquid-scintillator detector, the relevant signal observable is the total visible energy deposited by the final-state particles. We therefore construct visible energy $E_{\rm vis}$ event by event from the visible particles in the final state, including $\mu^\pm$, $\pi^\pm$, $p$, $e^\pm$, and $\gamma$, while not adding neutron-capture energy to $E_{\rm vis}$. 
Although JUNO can detect events down to visible energies of $\mathcal{O}(0.1)$ MeV, the atmospheric-neutrino background at low visible energies ($E_{\text{vis}} \lesssim 15$ MeV) suffers from large uncertainties~\cite{Honda:2015fha}. We therefore define our baseline sensitivity using a more conservative Large Energy Singles (LES)-inspired selection~\cite{Chauhan:2021fzu}, $15~\mathrm{MeV} < E_{\rm vis} < 100~\mathrm{MeV}$, together with a neutron veto.

The dominant background for energetic dark meson scattering in JUNO arises from neutral-current interactions of atmospheric neutrinos~\cite{Honda:2006qj,Honda:2011nf,Honda:2015fha,Diurba:2025lky}.
Following the general LES strategy of Ref.~\cite{Chauhan:2021fzu}, we calculate the background contribution from atmospheric neutrinos with $E_\nu > 100$ MeV using the flux from Ref.~\cite{Honda:2015fha}, and simulate neutral-current interactions on $^{1}$H and $^{12}$C with \textsc{GENIE}. For an exposure of $20~\mathrm{kton}\cdot\mathrm{year}$ in the signal window $15~\mathrm{MeV}<E_{\rm vis}<100~\mathrm{MeV}$, we obtain $N_{\rm bkg}^{\rm LES}=169.5$ events, of which $117.8$ originate from $^{12}$C and $51.7$ from $^{1}$H. We do not include the atmospheric-neutrino flux for $E_\nu<100~\mathrm{MeV}$ in our background estimation. 
As noted in Ref.~\cite{Chauhan:2021fzu}, this component carries an $\sim 25\%$ uncertainty and populates predominantly the visible-energy region below $15~\mathrm{MeV}$, so its contribution to the $E_{\text{vis}}\in(15,100)~\mathrm{MeV}$ window is expected to be marginal.

\begin{figure}[htbp]
    \centering
    \includegraphics[width=0.8\textwidth]{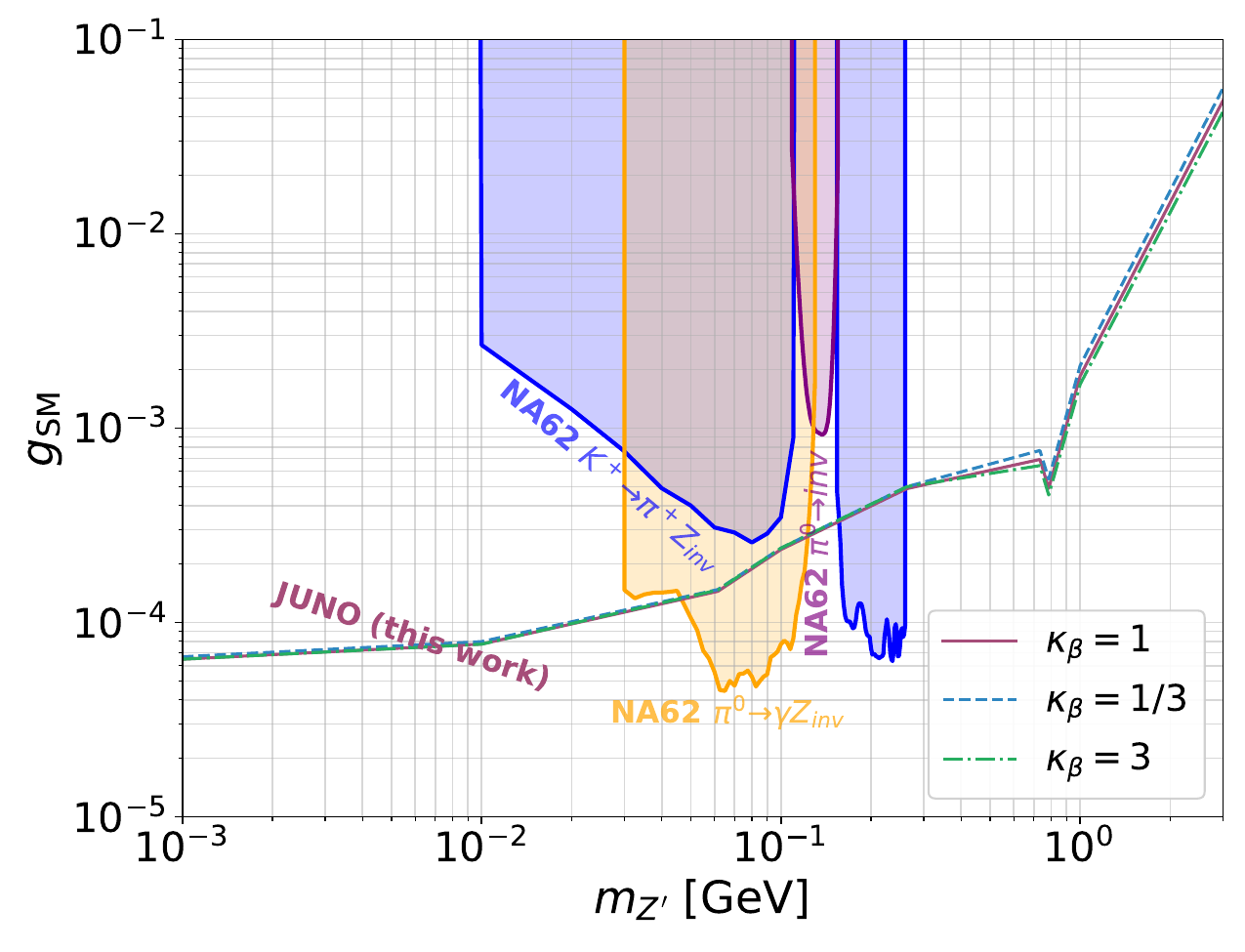}
    \caption{Projected 90\% C.L. sensitivity to the coupling $g_{\rm SM}$ as a function of the mediator mass $m_{Z^\prime}$ at JUNO, derived from the LES-inspired conservative event selection. The sensitivity is evaluated for an exposure of $20~\mathrm{kton}\!\cdot\!\mathrm{year}$. The solid, dashed, and dash-dotted curves correspond to the benchmark hadronization choices $\kappa_\beta=1$, $1/3$, and $3$, respectively. The shaded regions denote the parameter space excluded by existing NA62 constraints~\cite{NA62:2019meo,NA62:2025upx}.}
    \label{fig:gsm_sensitivity}
\end{figure}

In our model, the signal yield scales as $N_{\rm sig}(g_{\rm SM}) \propto g_{\rm SM}^{4}$, with one factor of $g_{\rm SM}^2$ from the partonic $Z'$ production and the other from the $K_D$–nucleon scattering cross section.
We quantify the projected sensitivity using the Asimov median significance~\cite{Cowan:2010js},
\begin{equation}
Z = \sqrt{ 2\left[ (N_{\rm sig}+N_{\rm bkg}) \ln \left(1+\frac{N_{\rm sig}}{N_{\rm bkg}}\right) - N_{\rm sig} \right] },
\label{eq:cowan_Z}
\end{equation}
and define the projected 90\% C.L. upper limit $g_{\rm SM}^{90}$ at each mediator mass by requiring $Z = 1.28$ with $N_{\rm bkg}=N_{\rm bkg}^{\rm LES}=169.5$. For this background, Eq.~\eqref{eq:cowan_Z} gives $N_{\rm sig}^{90} \simeq 17$. The sensitivity curves shown in Figure~\ref{fig:gsm_sensitivity} are then obtained directly from the LES-window signal yield at each scanned $m_{Z'}$, computed using the propagated dark-meson flux and the $K_D$–nucleus scattering cross sections.
\begin{table}[htbp]
\centering
\begin{minipage}{0.98\textwidth}
\centering
\renewcommand{\arraystretch}{1.25}
\setlength{\tabcolsep}{5pt}
\begin{threeparttable}
\begin{tabular*}{\textwidth}{@{\extracolsep{\fill}}ccccc@{}}
\toprule
$m_{Z^\prime}$~[GeV] & $N_{\rm incl}$ & $\epsilon_{n\text{-veto}}$ & $N_{\rm LES}$ & $g_{\rm SM}^{90}$ \\
\midrule
0.001 
& $157.482\,(156.377_{\rm C}+1.105_{\rm H})$ 
& 0.614 
& $96.693\,(95.619_{\rm C}+1.074_{\rm H})$ 
& $6.48\times10^{-5}$ \\

0.01  
& $75.787\,(75.341_{\rm C}+0.446_{\rm H})$ 
& 0.623 
& $47.213\,(46.791_{\rm C}+0.421_{\rm H})$ 
& $7.75\times10^{-5}$ \\

0.1   
& $0.840\,(0.758_{\rm C}+0.082_{\rm H})$ 
& 0.638 
& $0.537\,(0.456_{\rm C}+0.081_{\rm H})$ 
& $2.37\times10^{-4}$ \\

\bottomrule
\end{tabular*}
\caption{Expected JUNO signal yields and projected 90\% C.L. sensitivities for representative mediator masses in the visible-energy window $15~\mathrm{MeV}<E_{\rm vis}<100~\mathrm{MeV}$, evaluated at the reference coupling $g_{\rm SM}^{\rm ref}=10^{-4}$ for an exposure of $20~\mathrm{kton}\!\cdot\!\mathrm{year}$. Here $N_{\rm incl}$ denotes the inclusive signal yield obtained without imposing a final-state neutron veto, while $N_{\rm LES}$ denotes the LES-inspired conservative yield obtained after additionally vetoing events containing final-state neutrons. 
The event yields from carbon and hydrogen targets are shown separately in parentheses.
The ratio $\epsilon_{n\text{-veto}}\equiv N_{\rm LES}/N_{\rm incl}$ quantifies the corresponding signal retention under the neutron-veto requirement. 
\label{tab:juno_yields_les}}
\end{threeparttable}
\end{minipage}
\end{table}

To illustrate the size and structure of the signal yield entering the sensitivity evaluation, Table~\ref{tab:juno_yields_les} collects representative JUNO signal yields at three benchmark mediator masses in the visible-energy window $15~\mathrm{MeV}<E_{\rm vis}<100~\mathrm{MeV}$, evaluated at a reference coupling $g_{\rm SM}^{\rm ref}=10^{-4}$ for an exposure of $20~\mathrm{kton}\!\cdot\!\mathrm{year}$. It shows both the inclusive signal yield $N_{\rm incl}$ and the LES-inspired conservative yield $N_{\rm LES}$ obtained after vetoing events containing final-state neutrons. The ratio $\epsilon_{n\text{-veto}}\equiv N_{\rm LES}/N_{\rm incl}$ shows that this veto reduces the signal rate by roughly $40\%$ over the representative mass range considered here. As expected, the signal rate increases toward lighter mediator masses, reflecting the combined enhancement from the propagated flux and the scattering cross section. As a consistency check, we note that the signal retention ratio $\epsilon_{n\text{-veto}}\approx 0.6$ obtained here is broadly compatible with the $\sim 0.52$ proton-only fraction estimated in Ref.~\cite{Chauhan:2021fzu} for atmospheric-neutrino $\nu C$ QEL interactions. 
Moreover, the selected signal yield is found to be dominated by elastic scattering, whose contribution exceeds the inelastic one by roughly two orders of magnitude or more for these benchmark masses.

Figure~\ref{fig:gsm_sensitivity} shows the corresponding projected JUNO sensitivity in the $(m_{Z^\prime}, g_{\rm SM})$ plane for an exposure of $20~\mathrm{kton}\!\cdot\!\mathrm{year}$. The three curves correspond to the benchmark hadronization choices $\kappa_\beta=1$, $1/3$, and $3$, and therefore illustrate the residual theoretical uncertainty associated with the multiplicity prescription introduced in Section~\ref{sec:3}. Although the precise reach shifts moderately under this variation, the overall sensitivity pattern remains stable: JUNO probes couplings at or below the $10^{-4}$ level over much of the sub-$\mathcal{O}(0.1)~\mathrm{GeV}$ region and reaches the few$\times10^{-4}$ level around $m_{Z^\prime}\sim0.1~\mathrm{GeV}$. The shaded regions highlight existing exclusion limits from the NA62 experiment~\cite{NA62:2019meo,NA62:2025upx}. The upper mass boundary of the NA62 exclusion is determined kinematically by the mass difference between the kaon and pion, while the lower mass boundary is constrained primarily by significant missing-photon backgrounds in the low missing-invariant-mass region. We find that the qualitative complementarity with NA62 is unchanged under the $\kappa_\beta$ variation. JUNO remains sensitive in the low-mass window where NA62 rapidly loses coverage, and it stays competitive again once the NA62 reach is kinematically limited at larger mediator masses.
Astrophysical constraints, such as those arising from stellar cooling~\cite{An:2013yfc} and supernova bounds~\cite{Hardy:2024gwy}, could also apply to a light lepton-phobic mediator. However, given the significant theoretical uncertainties associated with these constraints and their strong dependence on the mediator's decay length and decay channels~\cite{Kazanas:2014mca,Sung:2019xie,DeRocco:2019njg,Caputo:2022mah,KA:2023dyz,Fiorillo:2025yzf}, they are not taken into account in this analysis.

\section{Conclusions} \label{sec:6} 
In this work, we have presented a comprehensive study on the atmospheric production and terrestrial detection of sub-GeV dark mesons within the framework of a confining dark sector coupled to the SM via a leptophobic $U(1)_D$ vector portal.  
By leveraging the high flux of cosmic rays and the large fiducial mass of the JUNO detector, we explored the sensitivity to light dark matter candidates which remain challenging to probe at collider experiments. 

In the literature, the transition from dark quarks to dark mesons is often implemented using phenomenological models that are difficult to extrapolate to confinement scales far from that of the SM QCD. To address this, we implemented a modified Quark Combination Model to describe the non-perturbative dark hadronization process. 
We generalized the existing QCM to accommodate vastly varying dark confinement scales.
In our setup, we observe that for a fixed kinematic ratio $\sqrt{s}/m_{q_k}$, a heavier dark sector scale yields higher meson multiplicities. 
The momenta of the final-state mesons are reconstructed via Monte-Carlo simulation utilizing the longitudinal phase space approximation.  

We evaluate the atmospheric dark meson flux arising from three distinct production mechanisms: proton bremsstrahlung, SM meson decay ($\pi^0,\eta \to \gamma Z^\prime$), and Drell-Yan processes. Our analysis of the flux composition demonstrates that radiative decays of light pseudoscalar mesons dominate the production in the low-mass mediator regime $m_{Z^\prime} \lesssim m_\eta$, providing a potent source of light dark matter. As for heavier mediators ($m_{Z^\prime} \gtrsim 1$ GeV), the Drell-Yan process becomes the primary production channel, despite the steep fall-off of the cosmic ray spectrum.

At detector level, we used the \textsc{GENIE} generator with a customized boosted-dark-matter module to simulate the scattering of relativistic dark mesons in the JUNO liquid-scintillator target. 
Both elastic and deep-inelastic scattering channels on carbon and hydrogen nuclei were included. 
The resulting event samples show that elastic scattering on carbon gives the dominant contribution over most of the parameter space considered here, while the hydrogen and DIS contributions are generally subleading after the visible-energy selection.

To address the atmospheric-neutrino background, we adopt a conservative LES-inspired signal definition for sensitivity estimation. 
We require
$15~\mathrm{MeV}<E_{\rm vis}<100~\mathrm{MeV}$ and veto events containing final-state neutrons. 
Using the atmospheric-neutrino flux in literature and applying the same event-level visible-energy construction as in the signal selection, we obtain an approximate background normalization
$N_{\rm bkg}^{\rm LES}=169.5$ for a $20~\mathrm{kton}\!\cdot\!\mathrm{year}$ exposure. 
We derived projected 90\% C.L. sensitivities to the coupling $g_{\rm SM}$ using the Asimov median significance. 
For the representative mediator masses $m_{Z^\prime}=0.001$, $0.01$, and $0.1~\mathrm{GeV}$, the corresponding projected reaches are
$g_{\rm SM}^{90}=6.48\times10^{-5}$, $7.75\times10^{-5}$, and $2.37\times10^{-4}$, respectively. 
We also examined the dependence on the phenomenological MQCM multiplicity parameter by rescaling
$\beta(\Lambda_D)\to\kappa_\beta\beta(\Lambda_D)$ with $\kappa_\beta=1/3,\,1,$ and $3$. 
The resulting sensitivity curves change only mildly under this order-one variation, indicating that the projected JUNO reach is not driven by a finely tuned choice of the benchmark hadronization parameter.

The JUNO sensitivity is complementary to existing NA62 constraints. 
In particular, JUNO can probe low mediator masses where accelerator searches are weakened by kinematic and background limitations, while the sensitivity at larger mediator masses is mainly controlled by the falling atmospheric dark-meson flux and by the reduced scattering rate. 
These results demonstrate that large-volume neutrino detectors can provide a useful complementary probe of sub-GeV dark-sector particles, even after adopting a conservative background-aware event selection.

\begin{acknowledgments}
This work was supported by the Natural Science Foundation of Sichuan Province under grant No. 2026NSFSC0034, by the National Natural Science Foundation of China (Project No. 11905149, 12505121), by the Joint Fund of Henan Province Science and Technology R$\&$D Program (Project No. 245200810077), by the Startup Research Fund of Henan Academy of Sciences (Project No. 20251820001), and by the Scientific and Technological Research Project of Henan Academy of Sciences (Project No. 20262320001).
\end{acknowledgments}

\bibliographystyle{jhep}
\bibliography{DarkMeson}

\end{document}